\documentclass[notitlepage,nofootinbib,tightenlines]{revtex4-2}
\usepackage{jltdefs}

\graphicspath{{figs/}}

\newcommand{\keep}[1]{}

\newcommand{\Vol}[1]{\lvert #1\rvert}

\renewmathnotation{\div}{\grad\cdot}

\mathnotation{\dV}{\dint{V}}
\mathnotation{\dS}{\dint{S}}
\mathnotation{\dSv}{\dint{\bm{S}}}

\renewcommand{\d}{\mathrm{d}}
\mathnotation{\dt}{\d{\t}}
\renewcommand{\t}{t}

\mathnotation[position vector]{\xv}{\bm{\xc}}
\mathnotationentry[domain]{\Omega}
\mathnotationentry[volume]{\Vol{\Omega}}
\mathnotationentry[domain boundary]{\pd\Omega}
\mathnotationentry[outward unit normal to~$\pd\Omega$]{\nuv}
\mathnotation[scalar concentration]{\cc}{\theta}
\mathnotation[probability density]{\pp}{p}

\mathnotation[scalar diffusivity]{\Dc}{D}
\mathnotation[diffusion tensor]{\Dbb}{\mathbb{\Dc}}

\mathnotation{\uc}{u}
\mathnotation[velocity]{\uv}{\bm{\uc}}
\mathnotation{\Fc}{F}
\mathnotationentry[fluid density]{\rho}
\mathnotation[flux vector]{\Fv}{\bm{\Fc}}
\mathnotation[number of particles]{\N}{N}
\mathnotation[Green's function]{\PP}{P}

\mathnotation[convex function]{\ffc}{f}
\newcommand{\fdiv}{$\ffc$-divergence}
\mathnotation{\fdivD}{H}
\mathnotation[\fdiv]{\Df}{\fdivD_{\ffc}}
\mathnotation{\uu}{u}
\mathnotation{\qq}{\pp_{12}}

\mathnotation[Kullback--Leibler divergence]{\DKL}{\fdivD_{\mathrm{KL}}}
\mathnotation[Jensen--Shannon divergence]{\DJS}{\fdivD_{\mathrm{JS}}}

\mathnotationentry
  [$L^\qn$ norm
   $\lVert\cdot\rVert_\qn = \l(\int_\Omega
   \lvert\cdot\rvert^\qn\dV\r)^{1/\qn}$,
   $1\le \qn\le\infty$]
  {\lVert\cdot\rVert_\qn}

\mathnotation[uniform velocity]{\Uc}{U}
\renewmathnotation[domain size]{\L}{L}

\mathnotation[number density]{\nn}{n}
\mathnotation[interior net source-sink]{\Q}{Q}
\renewmathnotation[interior source]{\S}{S}
\mathnotation[interior sink $-\K\nn$]{\K}{K}
\mathnotation[boundary net source-sink]{\q}{q}
\mathnotation[boundary sink $-\k\nn$]{\s}{s}
\renewmathnotation[boundary source]{\k}{k}

\mathnotation[$\hapf(\uu) \ldef (\uu - 1)\ffc'(\uu) - \ffc(\uu)$]%
  {\hapf}{g_{\ffc}}

\mathnotation{\qn}{q}

\begin{document}

\title{Nonuniform mixing}

\author{Jean-Luc Thiffeault}
\affiliation{Department of Mathematics, University of Wisconsin --
  Madison, 480 Lincoln Dr., Madison, WI 53706, USA}

\date{\today}

\begin{abstract}
  Fluid mixing usually involves the interplay between advection and diffusion,
  which together cause any initial distribution of passive scalar to
  homogenize and ultimately reach a uniform state.  However, this scenario
  only holds when the velocity field is nondivergent and has no normal
  component to the boundary.  If either condition is unmet, such as for active
  particles in a bounded region, floating particles, or for filters, then the
  ultimate state after a long time is not uniform, and may be time dependent.
  We show that in those cases of nonuniform mixing it is preferable to
  characterize the degree of mixing in terms of an \fdiv, which is a
  generalization of relative entropy, or to use the~$L^1$ norm.  Unlike
  concentration variance ($L^2$ norm), the \fdiv\ and~$L^1$ norm always decay
  monotonically, even for nonuniform mixing, which facilitates measuring the
  rate of mixing.  We show by an example that flows that mix well for the
  nonuniform case can be drastically different from efficient uniformly mixing
  flows.
\end{abstract}

\maketitle

\section{Introduction}
\label{sec:intro}

\subsection{Uniform mixing}

The standard paradigm for mixing in fluids is as follows
\cite{ThiffeaultAosta2004, Aref2017, GFD1999, Thiffeault2012, Doering2020}.
Initially, some passive scalar (such as red dye or virus particles) is
inhomogeneously distributed in a fluid.  Given enough time, the dye would
diffuse and spread uniformly throughout the domain; stirring the fluid greatly
enhances the speed of this homogenization process.  The ultimate steady state
is a fluid with uniform concentration of dye throughout the domain.

The mathematical underpinning for this process is straightforward.  The dye
concentration~$\cc(\xv,\t)$ obeys the advection-diffusion equation
\begin{equation}
  \pd_\t\cc + \uv\cdot\grad\cc = \Dc\,\grad^2\cc
  \label{eq:AD}
\end{equation}
where the velocity field~$\uv(\xv,\t)$ is nondivergent ($\div\uv=0$),
and~$\Dc>0$ is the dye diffusivity.  Since a constant~$\cc$
solves~\eqref{eq:AD}, we can assume without loss of generality
that~$\int_\Omega\cc\dV=0$, that is, $\cc$ has zero mean over the bounded
domain~$\Omega$.  In that case we find after a few integrations by parts
\begin{equation}
  \frac{\d}{\dt}\int_\Omega \cc^2\dV
  =
  -2\Dc\int_\Omega\lvert\grad\cc\rvert^2\dV
  \le 0,
  \label{eq:ADvar}
\end{equation}
where boundary terms vanish, assuming no-flux boundary conditions
on~$\cc$. \Cref{eq:ADvar} gives the evolution of the \emph{concentration
  variance} or~$L^2$ norm of~$\cc$, and the nonpositivity of the right-hand
side shows that variance will decrease until~$\cc$ is a constant throughout
the whole domain~$\Omega$.  This constant vanishes because of the zero-mean
assumption, so the ultimate steady state is~$\theta\equiv0$ everywhere.  We
then declare the dye to be mixed.  This argument makes no reference to the
velocity~$\uv(\xv,\t)$, since the terms involving it have integrated away.
\Cref{eq:ADvar} thus cannot be used to get a useful estimate of the rate of
mixing.  Nevertheless, simply having an equation such as~\eqref{eq:ADvar} is
essential in mathematical analysis since it guarantees mixing for long enough
times, no matter what the form of~$\uv$.  It also validates the common use of
variance as a measure of the degree of mixing.  The right-hand side of
\cref{eq:ADvar} is called the \emph{variance dissipation}, and the magnitude
of its integrand is a useful proxy for regions where mixing is most active.

\begin{figure}
  \begin{center}
    \includegraphics[width=.4\textwidth]{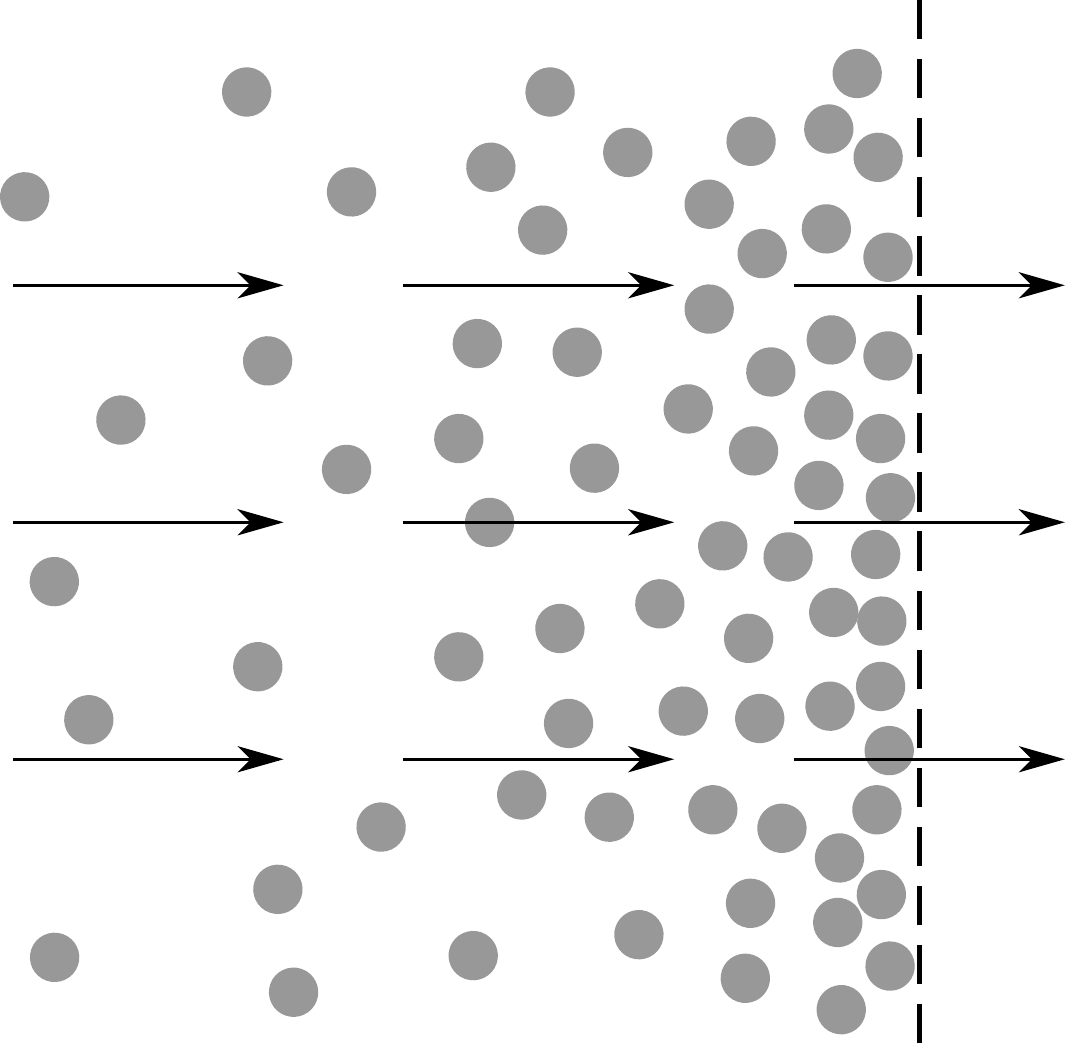}
  \end{center}
  \caption{In the presence of a suction flow, passive particles such as
    viruses accumulate near a filter.  The filter is permeable to fluid but
    not to particles.}
  \label{fig:particles_filter}
\end{figure}

\subsection{Compressibility}

When the fluid is compressible, the fluid density~$\rho(\xv,\t) > 0$ is solved
for along with the concentration~$\cc(\xv,\t)$, and instead of~\eqref{eq:AD}
we have the coupled equations
\begin{equation}
  \pd_\t\rho + \div(\uv\rho) = 0,
  \qquad
  \pd_\t(\rho\,\cc) + \div(\uv\rho\,\cc) = \Dc\,\grad^2\cc.
  \label{eq:ADcomp}
\end{equation}
Notice that~$\theta = \text{const.}$ is still a solution of~\eqref{eq:ADcomp},
so the ultimate steady state remains uniform.  The concentration variance
equation~\eqref{eq:ADvar} becomes
\begin{equation}
  \frac{\d}{\dt}\int_\Omega \rho\,\cc^2\dV
  =
  -2\Dc\int_\Omega\lvert\grad\cc\rvert^2\dV
  \le 0,
  \label{eq:ADvarcomp}
\end{equation}
again assuming no-flux boundary conditions on~$\theta$.  The variance will
relax to zero over time, implying that~$\theta(\xv,\t)$ reaches the uniform
mixed state.  In that sense compressible mixing is also an instance of a
uniform mixing scenario.

Note that setting~$\rho = \text{const.}$ in~\eqref{eq:ADcomp} necessarily
implies that~$\div\uv=0$.  Starting in the next section we shall allow for
cases where~$\div\uv \ne 0$, but where fluid density does not enter the
problem.  These cases are not the same as compressible mixing; we shall
usually refer to them as \emph{divergent flows} to avoid confusion.

\subsection{Nonuniform mixing}
\label{sec:nonuniform_mixing}

There was an implicit assumption when we stated that~$\cc = \text{const.}$ is
a steady solution of~\eqref{eq:AD}: we required~$\uv\cdot\nuv=0$ at the
boundary~$\pd\Omega$.  Furthermore, when \hbox{$\div\uv\ne0$} we must
modify~\eqref{eq:AD} to read
\begin{equation}
  \pd_\t\cc + \div(\uv\,\cc) = \Dc\,\grad^2\cc
  \label{eq:AD2}
\end{equation}
to ensure that~$\int_\Omega\cc\dV$ is conserved under no-flux boundary
conditions.  For~$\cc = \text{const.}$ to be a steady solution
of~\cref{eq:AD} or~\eqref{eq:AD2}, we require both~$\uv\cdot\nuv=0$ at the
boundary~$\pd\Omega$, as well as the nondivergence condition~$\div\uv=0$.  If
either of these conditions is not satisfied, then the steady state is not
uniform in space.  In fact there may even be no steady state at all, in which
case we instead refer to an \emph{ultimate state}, which is reached after a
long time.  We will define this ultimate state more precisely later.

The no-penetration condition~$\uv\cdot\nuv=0$ is usually quite reasonable: it
says that fluid doesn't go through the walls.  But in many relevant
applications the fluid can go through boundaries, even if the passive scalar
cannot.  We give two examples of such a situation.  (Note that we will use the
term `passive scalar' and `particles' somewhat interchangeably.  We usually
denote by~$\theta$ a passive scalar that can have either sign, and by~$\pp$
or~$\nn$ a particle density that cannot be negative.)

\paragraph*{\textbf{Particle filters.}}
If the fluid is air and the passive scalar consists of virus particles, then a
filter is a membrane that allows the passage of air but not of viruses
(hopefully).  This is shown schematically in \cref{fig:particles_filter}: the
virus particles naturally accumulate at the filter where~$\uv\cdot\nuv>0$ due
to the suction effect.  In this type of situation the `mixed state' is no
longer uniform because of this accumulation.

\paragraph*{\textbf{Active particles.}}
A popular model for 2D self-propelled active particles (so-called Janus
particles~\cite{Golestanian2007}) assumes that the particles move at a
constant speed~$U$, in a swimming direction given by an angle~$\phi$ that
evolves randomly with time \cite{vanTeeffelen2008, Kurtzhaler2016}.  The
probability density of particles~$\pp(\xv,\phi,\t)$ obeys a Fokker--Planck
(or Smoluchowski) equation
\begin{equation}
  \pd_\t\pp + (\uv + U\hat{\bm{q}})\cdot\grad\pp = \Dc\,\grad^2p +
  \Dc_{\mathrm{rot}}\,\pd_\phi^2\pp,
  \label{eq:FP_active}
\end{equation}
with~$\hat{\bm{q}} = (\cos\phi,\sin\phi)$ and rotational
diffusion~$\Dc_{\mathrm{rot}}$.  \Cref{eq:FP_active} is exactly analofous
to~\eqref{eq:AD}, except that the domain~$\Omega$ involves spatial
coordinates~$\xv$ and the angle~$\phi$. The fluid velocity~$\uv(\xv,\t)$
obeys~$\uv\cdot\nuv=0$ at boundaries, but the swimming
velocity~$U\hat{\bm{q}}$ does not: a particle may keep pushing against a
boundary even after it makes contact.  (It is prevented from entering the wall
by the no-flux boundary condition on~$\pp$.)  Hence, the steady solution to
\cref{eq:FP_active} is not uniform: particles tend to accumulate near
boundaries, in a manner similar to the filter example above \cite{Lee2013,
  Ezhilan2015, ChenThiffeault2021}.

There are two other effects that can lead to nonuniform ultimate states:
divergence of the velocity ($\div\uv\ne0$) and the presence of sources and
sinks.  We give examples for each case.

\paragraph*{\textbf{Divergent velocity.}}
Floating particles at the surface of the ocean are subjected to the fluid
velocity field~$\uv(\xv,\t)$ evaluated at~$\zc=0$.  Even though the
three-dimensional velocity satisfies~$\div\uv=0$, the two-dimensional velocity
at the surface is in general divergent.  In the long-time limit, particles
will tend to congregate at downwellings, where the divergence is negative.
The ultimate state is thus nonuniform~\cite{DAsaro2018}.

\mathnotation{\surf}{{\mathrm{s}}}

The same type of model applies to surfactants, which are concentration scalar
fields defined at the surface of a fluid.  The equation for a surfactant
concentration~$\theta(\xv,\t)$ evaluated at a free surface
is~\cite{Aris,Stone1990}
\begin{equation}
  \pd_\t\cc + \grad_\surf\cdot(\uv_\surf\,\cc)
  =
  \Dc\,\grad_\surf^2\cc
  - \cc\,(\grad_\surf\cdot\nuv_\surf)\,\uv\cdot\nuv_\surf
  \label{eq:surfactant}
\end{equation}
where~$\grad_\surf$ is a gradient along the surface, $\uv_\surf$ is the
component of~$\uv$ parallel to the surface, and~$\nuv_\surf$ is the unit
normal to the surface.  The source-sink term on the right vanishes if the
surface is flat~$(\grad_\surf\cdot\nuv_\surf=0)$ or if it is not
moving~$(\uv\cdot\nuv_\surf=0)$.  Even though the three-dimensional velocity
is nondivergent, the surface divergence~$\grad_\surf\cdot\uv_\surf$ is
generally nonzero.  \Cref{eq:surfactant} thus has the form of \cref{eq:AD2},
and the surfactant concentration can achieve a nonuniform ultimate state.

\paragraph*{\textbf{Heating a room.}}
In the winter, a closed room may be heated by a space heater, which is a
localized source of heat.  A closed window somewhere else in the room may act
as a sink of heat.  The equilibrium state is nonuniform: after a long time, we
still expect the temperature to be warmer near the heater, and cooler near the
window.

Whenever \cref{eq:AD2} fails to have a uniform steady state, we are dealing
with \emph{nonuniform mixing}: any initial condition~$\theta(\xv,\t_0)$ still
tends towards an ultimate state, and stirring can accelerate this convergence.
However, mixing must be defined with respect to this ultimate state, not the
uniform state.  Note that this ultimate state may be time-dependent, which
challenges our natural notion of mixing even further.

Note that even if~$\uv\cdot\nuv\ne0$ on the boundary~$\pd\Omega$, it is still
typically the case that
\begin{equation}
  \int_\Omega \rho\,\uv\cdot\nuv\dS = 0,
  \label{eq:mass_cons}
\end{equation}
where~$\rho$ is the fluid density.  \Cref{eq:mass_cons} is a consequence of
fluid mass conservation inside~$\Omega$.  However, we shall not assume
that~\cref{eq:mass_cons} is satisfied in our development, since it is
unnecessary, and there are cases where fluid mass might not be conserved (for
instance, if there is some external source of fluid, such as rain).

\subsection{Convergence to the ultimate state}

When the variance equation~\eqref{eq:ADvar} is modified to allow for a
nonuniform ultimate state, as described in \cref{sec:nonuniform_mixing}, we
will see that it no longer implies monotonic convergence to that state,
because of a nonvanishing term that is sign-indefinite.  Concentration
variance becomes an unreliable measure of mixing, at least from a mathematical
viewpoint.

We will show that, in all cases where the ultimate state is nonuniform, the
degree of mixing is better captured by a kind of entropy function, related to
the relative entropy of information theory and statistical physics.  This
entropy function has a time evolution that is always nonincreasing, no matter
the subtleties of the system, and therefore always predicts convergence to an
ultimate state.

We also show that the~$L^1$ norm of~$\cc$ satisfies
\begin{equation}
  \frac{\d}{\dt}\int_\Omega \lvert\cc\rvert\dV
  =
  -2\Dc\int_{\{\cc=0\}}\lvert\grad\cc\rvert\dS
  \le 0
  \label{eq:ADL1}
\end{equation}
where the integral on the right is taken over the zero level set
of~$\cc(\cdot,\t)$.  \Cref{eq:ADL1} holds in the general case, unlike the
variance equation~\eqref{eq:ADvar} which depends on the
nondivergence~$\div\uv=0$ and~$\uv\cdot\nuv=0$ at the boundary~$\pd\Omega$.
Thus, in general the~$L^1$ norm is preferable to the~$L^2$ norm as a measure
of mixing, as we will make evident by simple numerical examples.  In fact we
will show that~$L^1$ is the only $L^\qn$ norm (with~$1 \le \qn \le \infty$)
having this monotonic decay property.

The main point of this article is that the nondivergence
condition~$\div\uv=0$ and no-penetration condition~$\uv\cdot\nuv=0$ lead to a
very special situation in that the ultimate mixed state is uniform.  This is
not true if either of these conditions is violated; we categorize the
resulting situations as nonuniform mixing.  We must then revise what we mean
by the rate of mixing: instead of defining it as the rate of approach to a
uniform state, it is preferable to use the rate at which any two initial
states converge to each other.

Nonuniform mixing can be very different from traditional mixing.  For example,
we will show by an example that a constant flow can be an exceedingly good
mixer in the presence of suction boundary conditions, whereas such a flow is
essentially useless for traditional mixing.  The reason is that with the
suction conditions the flow presses particles against one wall, which leads to
a rapid convergence of any two initial conditions towards each other.

The presentation in this paper is unapologetically mathematical: the aim is to
give the precise underpinnings in (hopefully) an agreeable language,
without the rigorous burden of function spaces.  In addition, the approach
presented here relies on some techniques common in the analysis of convergence
in Fokker--Planck equations \cite{Pavliotis, Riksen, Arnold2008,
  Achleitner2015, Arnold2018, Arnold2001, Lelievre2013}, which are standard in
statistical physics but less well known in fluid dynamics, even though the
mathematical framework is similar.  One major difference is that in
statistical physics one is typically less concerned with specific boundary
conditions, since the independent variables are often quantities like momenta,
which live in unbounded spaces.  In contrast, here we shall pay particularly
close attention to the role of boundary conditions.  Another difference is
that much of the Fokker--Planck literature involves cases where the steady
state exists and is easily identified, which will not be the case here for our
more complicated, time-dependent examples.

It is worth noting that entropies have been used by several authors to
quantify fluid mixing; see for instance \cite{DAlessandro1999, Stremler2006,
  Camesasca2006, Krutzmann2008, Fodor2011, Lauritzen2011, Grahn2012,
  Brandani2013, Perugini2015}.  However, their approaches are usually based on
measuring statistical properties, whereas here we focus directly on
differential equations to get rigorous bounds.  Entropy and mixing are also
often studied in the context of nonequilibrium thermodynamics, but in that
case there is usually an equilibrium state such as a Maxwell--Boltzmann
distribution towards which the system is tending.  Our description will be
more general and adapted to the context of fluid mixing.  Approaches based on
topological entropy~\cite{Boyland2000, Thiffeault2006, Gouillart2006} are
complementary but not closely related to ours, since they focus on properties
of trajectories of~$\uv(\xv,\t)$.

Our paper is organized as follows.  In \cref{sec:closed} we define the system
and derive some basic results.  We consider the simplest `traditional' case of
nondivergent flow with impermeable boundary conditions in
\cref{sec:incomp_imperm}, and show that the time-evolution equation for
variance in that case predicts convergence to a uniform state.  In
\cref{sec:comp_or_perm} we relax both the nondivergence and impermeability
conditions.  Now the ultimate state is no longer uniform, and may not even be
steady.  The variance equation no longer implies convergence, due to the
addition of a sign-indefinite term.  We remedy this by introducing the \fdiv\
associated with two probability densities~$\pp_1$ and~$\pp_2$, a quantity that
arises in information theory.  (A special case of the~\fdiv\ is the relative
entropy of~$\pp_1$ and~$\pp_2$.)  We show that the time evolution of the
\fdiv\ is nondecreasing, and that it must eventually decrease to zero.

We give some simple examples for flows that can be fully solved in
\cref{sec:examples}.  In particular, we show that a constant flow with suction
boundary conditions can be surprisingly effective at mixing.  In
\cref{sec:sources} we incorporate the effect of sources and sinks.  For those
we need to slightly generalize the definition of~\fdiv, and we can still show
convergence to an ultimate state.  We discuss the time evolution of the~$L^1$
norm in \cref{sec:TV}.  Finally, we offer some concluding remarks in
\cref{sec:disc}.

\section{A particle in a closed domain}
\label{sec:closed}

Consider a particle in a closed, connected domain~$\Omega$.  The particle
could represent a virus, or some molecule of a pollutant.  The particle
evolves according to a velocity field (or drift) $\uv(\xv,\t)$ and a diffusion
tensor~$\Dbb(\xv,\t)$.  The probability of finding the particle in a small
volume~$\dV_{\xv}$ centered on~$\xv$ is~$\pp(\xv,\t)\dV_{\xv}$, where the
probability density obeys the Fokker--Planck equation
\begin{equation}
  \pd_\t\pp + \div\Fv(\pp) = 0,
  \qquad
  \xv \in \Omega,
  \label{eq:FP}
\end{equation}
with the probability flux defined as
\begin{equation}
  \Fv(\pp) \ldef \uv(\xv,\t)\,\pp - \Dbb(\xv,\t)\cdot\grad\pp\,.
  \label{eq:Fv}
\end{equation}
The probability flux consists of an advective part and a diffusive part.  In
the fluid-dynamical context \cref{eq:FP} is called an advection-diffusion
equation.

The probability density satisfies~$\pp\ge0$ and~$\int_\Omega\pp\dV \le 1$.
(If the total probability is less than one, then the particle might not be in
the domain at all.)  We can integrate \cref{eq:FP} over~$\Omega$ and use the
divergence theorem to get
\begin{equation}
  \frac{\d}{\dt}\int_\Omega \pp(\xv,\t)\dV
  =
  -\int_{\pd\Omega} \Fv(\pp)\cdot\dSv\,,
  \label{eq:dt_intpp}
\end{equation}
where~$\dSv = \nuv\dS$, with~$\nuv$ the outward unit normal to the
boundary~$\pd\Omega$.  \Cref{eq:dt_intpp} makes it clear that we can conserve
total probability by requiring the no-flux boundary condition
\begin{equation}
  \Fv(\pp)\cdot\nuv = 0,
  \qquad
  \xv \in \pd\Omega.
  \label{eq:noflux}
\end{equation}
It is important to note that we have not made any assumptions
on~$\uv(\xv,\t)$, other than a bit of smoothness.  In particular we did not
assume~$\div\uv=0$.  In addition, we did not assume~$\uv\cdot\nuv=0$, so the
boundary condition \cref{eq:noflux} is of mixed type (i.e., a linear
combination of~$\pp$ and~$\grad\pp$).

We spoke of one particle in this section, but the description works equally
well for~$\N$ noninteracting particles, with~$\N$ fixed, or if~$\pp$ is a
nonnegative quantity such as heat, appropriately normalized.  Later in
\cref{sec:sources}, we will introduce sources and sinks, so that~$\N$ will be
allowed to vary.

\section{Nondivergent flow with impermeable boundary}
\label{sec:incomp_imperm}

We make the additional assumptions
\begin{subequations}
\label{eq:incomp_imperm}
\begin{alignat}{2}
  \div\uv = 0, \quad &\xv\in\Omega,
  \label{eq:incomp}\\
  \qquad
  \uv\cdot\nuv = 0, \quad &\xv\in\pd\Omega.
\label{eq:imperm}
\end{alignat}
\end{subequations}
\Cref{eq:incomp} is the \emph{nondivergence condition}, and
\cref{eq:imperm} is the \emph{impermeability condition}.  The no-flux boundary
condition~\eqref{eq:noflux} reduces to~$\nuv\cdot\Dbb\cdot\grad\pp=0$.

Observe that, under conditions~\eqref{eq:incomp_imperm}, \cref{eq:FP} with
boundary conditions~\eqref{eq:noflux} has the steady solution~$\pp = \varphi$,
with
\begin{equation}
  \varphi(\xv) = {\Vol{\Omega}}^{-1},
  \label{eq:uniform}
\end{equation}
where~$\Vol{\Omega}$ is the volume of~$\Omega$, so
that~$\int_\Omega\varphi\dV = 1$.  The
solution~$\varphi(\xv) = {\Vol{\Omega}}^{-1}$ is called the uniform density
on~$\Omega$.  Two important remarks are in order: (i) \emph{both} conditions
in~\eqref{eq:incomp_imperm} are necessary for \cref{eq:uniform} to be a steady
solution; (ii) \cref{eq:uniform} is a steady solution even when~$\uv(\xv,\t)$
and~$\Dbb(\xv,\t)$ are explicit functions of time.

We define mixing as the tendency for any initial condition~$\pp(\xv,\t_0)$ to
converge to~$\varphi(\xv)$ as~$\t \rightarrow \infty$.  A traditional way of
characterizing this convergence is to first define the anomaly
\begin{equation}
  \theta(\xv,\t) \ldef \pp(\xv,\t) - \varphi(\xv)
  \label{eq:anomaly}
\end{equation}
so that~$\int_\Omega \theta\dV=0$.  The \emph{variance} is
then~$\int_\Omega \theta^2\dV$; after a few integrations by parts, we find
that it evolves according to
\begin{equation}
  \frac{\d}{\dt}\int_\Omega\theta^2 \dV
  =
  \int_\Omega
  \uv\cdot\grad\theta^2\dV
  -
  2\int_\Omega\grad\theta\cdot\Dbb\cdot\grad\theta\dV.
  \label{eq:var_evol}
\end{equation}
The first integral on the right vanishes:
from~\eqref{eq:incomp}~$\uv\cdot\grad\theta^2 = \div(\uv\theta^2)$, followed
by the divergence theorem and then~\eqref{eq:imperm}.  Next we require that
there exists a constant~$\sigma>0$ such that
\begin{equation}
  \bm{v}\cdot\Dbb(\xv,\t)\cdot\bm{v}
  \ge \sigma\lvert\bm{v}\rvert^2 > 0,
  \qquad
  \text{for all nonzero vectors $\bm{v}$},
  \label{eq:uniform_elliptic}
\end{equation}
i.e., the operator~$\div(\Dbb\cdot\grad\pp)$ is \emph{uniformly elliptic}.
With~\eqref{eq:uniform_elliptic}, \cref{eq:var_evol} now gives
\begin{equation}
  \frac{\d}{\dt}\int_\Omega\theta^2 \dV
  \le
  -
  2\sigma\int_\Omega\lvert\grad\theta\rvert^2\dV
  \le
  -
  2\sigma\lambda\int_\Omega\theta^2\dV
  \label{eq:var_evol_bound}
\end{equation}
where in the last step we used the Poincar\'e--Wirtinger
inequality~$\lVert \grad\theta\rVert_2^2 \ge \lambda\,\lVert\theta\rVert_2^2$
for a mean-zero function~$\theta$~%
\footnote{The $L^\qn$ norm $\lVert f\rVert_\qn$ is defined by
  $\lVert f\rVert_\qn = \l( \int_\Omega \lvert f\rvert^\qn\dV\r)^{1/\qn}$
  for~$1\le \qn\le\infty$.}.  The constant~$\lambda>0$ depends only on the
domain~$\Omega$.  Gr\"{o}nwall's lemma then yields the bound
\begin{equation}
  \int_\Omega\theta^2(\xv,\t) \dV
  \le
  \ee^{-2\sigma\lambda(\t-\t_0)}\int_\Omega\theta^2(\xv,\t_0) \dV
  \label{eq:var_conv}
\end{equation}
which goes to zero as~$\t\rightarrow\infty$.  We conclude
that~$\theta \rightarrow 0$, or~$\pp \rightarrow \varphi$.  Thus the ultimate
fate of any initial~$\pp(\xv,\t_0)$ is to be homogenized until the probability
of finding the particle anywhere in~$\Omega$ is uniform.  The rate at which
this happens is of order~$2\sigma\lambda$, though this is generally an
underestimate.  In practice, the action of~$\uv(\xv,\t)$, called
\emph{stirring}, amplifies gradients so that~$\grad\theta$ in
\cref{eq:var_evol} can be much larger than required by the
Poincar\'e--Wirtinger inequality.  Nevertheless, \cref{eq:var_conv} is useful
in that it proves that variance \emph{must} converge to zero.  What we have
just described is the basic idea of what is traditionally meant by mixing in
the fluids community.

What can happen if we violate the uniform ellipticity condition
\cref{eq:uniform_elliptic}?  For example, consider the heat equation
\begin{equation}
  \pd_\t\pp = T'(\t)\,\grad^2\pp
\end{equation}
with time-dependent diffusion coefficient~$\Dc(\t) = T'(t)$.
If~$T'(t) \sim t^{-\alpha}$ for large time, then the uniform ellipticity
condition is violated when~$\alpha>0$.  We can rescale and use~$T$ as a time
coordinate, in which case we expect a long-time exponential decay of the form
\begin{equation}
  \pp(\xv,\t) - \Vol{\Omega}^{-1}
  \sim \ee^{-\gamma T(\t)}\,,
\end{equation}
where~$\gamma$ is the asymptotic decay rate for~$T'(\t)=1$.
Since~$T(t) \sim t^{-\alpha+1}$, we see that~$\pp$ will fail to converge to
the uniform density for~$\alpha>1$.  Thus, the
condition~\eqref{eq:uniform_elliptic} is only sufficient: there may still be
convergence to equilibrium even if it is not satisfied.

\section{Divergent flow or permeable boundary}
\label{sec:comp_or_perm}

The situation described in \cref{sec:incomp_imperm} is straightforward: for
any velocity field~$\uv(\xv,\t)$ and diffusion tensor~$\Dbb(\xv,\t)$, we can
expect convergence to a uniform density as long as
conditions~\eqref{eq:incomp_imperm} and~\eqref{eq:uniform_elliptic} are
satisfied.  Now we investigate what happens when either the flow is
divergent (\cref{eq:incomp} not satisfied), or when there is suction of
fluid through the boundary (\cref{eq:imperm} not satisfied).

First consider the autonomous case where~$\uv(\xv,\t) \rightarrow \uv(\xv)$,
and $\Dbb(\xv,\t) \rightarrow \Dbb(\xv)$.  Then there is an equilibrium
density~$\varphi(\xv) >0$ that satisfies
\begin{equation}
  \grad\cdot(\uv\,\varphi - \Dbb\cdot\grad\varphi) = 0,
  \quad \xv \in \Omega;
  \qquad
  \Fv(\varphi)\cdot\nuv = 0, \quad \xv \in \pd\Omega
\end{equation}
and is normalized: $\int_\Omega\varphi\dV=1$.  We can then define the anomaly
as we did in \cref{eq:anomaly}; the only difference is that the reference
state~$\varphi(\xv)$ is no longer uniform.  The variance evolution
equation~\eqref{eq:var_evol} is still valid, but now the first integral term
on the right now longer vanishes.  This term is not sign-definite: this means
that we can no longer conclude from this equation alone that variance must
decay.  In fact, variance \emph{does} eventually decay, but it might not do so
monotonically.  \Cref{eq:var_evol} alone is not enough to conclude that~$\pp$
converges to~$\varphi$.

It would be convenient, then, to have a quantity other than variance that does
decay monotonically in this general case.  To that end, consider
the 
\fdiv\ of two normalized probability densities~$\pp_1(\xv)$
and~$\pp_2(\xv)$~\cite{Osterreicher2003, Liese2006}:
\begin{equation}
  \Df[\pp_1,\pp_2] \ldef
  \int_{\Omega}
  \pp_2\,\ffc(\pp_1/\pp_2)\dV.
  \label{eq:fdiv}
\end{equation}
Here~$\ffc: \mathbb{R}_{\ge0} \rightarrow \mathbb{R}$ is an arbitrary convex
function with~$\ffc(1)=0$.  The \fdiv\ is nonnegative; indeed, since~$\pp_2$
is a probability density, by Jensen's inequality for the convex
function~$\ffc$ we have
\begin{equation}
  \Df[\pp_1,\pp_2] \ge
  \ffc\l(\int_\Omega (\pp_1/\pp_2)\,\pp_2 \dV\r) = \ffc(1) = 0.
  \label{eq:Df_pos}
\end{equation}
The~\fdiv\ is zero if and only if~$\pp_1\equiv\pp_2$; $\Df[\pp_1,\pp_2]$
measures how different~$\pp_1$ and~$\pp_2$ are from each other---hence the
name `divergence.'  The \fdiv\ is not generally a metric for probability
densities, since~$\Df[\pp_1,\pp_2] \ne \Df[\pp_2,\pp_1]$, though for certain
choices of~$\ffc$ it can be made symmetric (see below).

We now further assume that~$\ffc$ is strictly convex and twice-differentiable.
If each~$\pp_i(\xv,\t)$ evolves according to \cref{eq:FP}, with no-flux
boundary condition~\eqref{eq:noflux}, then we show in \cref{apx:fdiv_dot} that
\begin{equation}
  \skew{-3}\dot\Df[\pp_1,\pp_2]
  =
  -
  \int_{\Omega}
  \pp_2\,\ffc''(\pp_1/\pp_2)\,
  \grad(\pp_1/\pp_2)\cdot\Dbb\cdot\grad(\pp_1/\pp_2)
  \dV
  \le 0
  \label{eq:fdiv_dot}
\end{equation}
since~$\ffc'' > 0$ for a strictly convex function.  For~$\Dbb$
satisfying~\eqref{eq:uniform_elliptic}, notice that the right-hand side
of~\eqref{eq:fdiv_dot} is zero if and only if~$\pp_1\equiv\pp_2$.  Hence, any
two solutions to~$\pd_\t\pp = -\grad\cdot\Fv(\pp)$ converge to each other; in
the autonomous case they converge to the fixed point~$\pp=\varphi$.

We emphasize that \cref{eq:fdiv_dot} holds for any divergent flow, possibly
with suction boundary conditions, with time-dependent~$\uv$ and~$\Dbb$.  In
that sense the \fdiv\ is a better descriptor of mixing than variance: it
monotonically decreases for any flow.  The evolution
equation~\eqref{eq:fdiv_dot} also suggests how to define mixing in the
nonautonomous context: $\pp_1$ and $\pp_2$ converge to some ultimate
state~$\varphi(\xv,\t)$, which is `locked' to the time-dependence of~$\uv$
and~$\Dbb$.  Thus, the main characteristic of mixing is not that it leads to a
homogeneous state, but rather that it leads to a state that has completely
forgotten the initial condition.  This ultimate state must be unique (for
connected $\Omega$), otherwise~\eqref{eq:fdiv_dot} leads to a contradiction.
Unfortunately, extracting an explicit bound on the decay rate
from~\eqref{eq:fdiv_dot} is much more challenging than it was in the case of
variance in \cref{eq:var_conv}, and is still a topic of ongoing research
\cite{Arnold2008, Achleitner2015, Arnold2018, Arnold2001}.

The discussion of~$\Df$ so far did not depend on a choice of the convex
function~$\ffc$ in~\eqref{eq:fdiv}, as long as it exists.  A simple choice
for~$\ffc$ is
\begin{equation}
  \ffc(\uu) = \uu\log \uu,
  \qquad
  \ffc''(\uu) = 1/\uu,
  \label{eq:ffc_KL}
\end{equation}
which corresponds to the \emph{relative entropy} or \emph{Kullback--Leibler
  divergence} (KLD), denoted by $\DKL(\pp_1,\pp_2)$~\cite{CoverThomas}~%
\footnote{These entropies tend to decrease to zero with time, which is the
  opposite definition to that used in physics.}:
\begin{equation}
  \DKL(\pp_1,\pp_2)
  =
  \int_\Omega
  \pp_1\log\l(\pp_1/\pp_2\r)
  \dV.
  \label{eq:KLD}
\end{equation}
The KLD can be interpreted as the amount of information lost
when~$\pp_2$ is used to approximate~$\pp_1$.  The KLD bounds the~$L^1$ norm by
\emph{Pinsker's inequality}: \keep{More general: Csiszr--Kullback inequality,
  see Arnold et al.  Is (2.1c) needed?}
\begin{equation}
  \lVert\pp_1 - \pp_2\rVert_1^2 \le 2\log2\,\DKL(\pp_1,\pp_2).
\end{equation}
However, $\DKL(\pp_1,\pp_2)$ is not symmetric in~$\pp_1$ and~$\pp_2$, and is
unbounded when~$\pp_2$ vanishes anywhere in~$\Omega$.  (We will discuss the
time evolution of~$\lVert\pp_1 - \pp_2\rVert_1$ in \cref{sec:TV}.)

A slightly more involved choice for~$\ffc$ is
\begin{equation}
  \ffc(\uu) = \tfrac12\uu\log\uu
  - \tfrac12(1 + \uu) \log[\tfrac12(1 + \uu)],
  \qquad
  \ffc''(\uu) = (2\uu\,(1 + \uu))^{-1},
  \label{eq:ffc_JS}
\end{equation}
which leads to the
{\emph{Jensen--Shannon divergence}} (JSD), denoted
by~$\DJS(\pp_1,\pp_2)$~\cite{Endres2003}:
\begin{align}
  \DJS(\pp_1,\pp_2)
  &=
  \tfrac12\l\{
  \DKL(\pp_1,\qq)
  +
  \DKL(\pp_2,\qq)
  \r\} \nonumber\\
  &=
  \tfrac{1}{2}
  \int_\Omega
  \l\{
  \pp_1\log\l(\pp_1/\qq\r)
  +
  \pp_2\log\l(\pp_2/\qq\r)
  \r\}
  \dV
  \label{eq:JSD}
\end{align}
where~$\qq \ldef \tfrac12(\pp_1 + \pp_2)$.  The JSD is symmetric in~$\pp_1$
and~$\pp_2$, and its square root is a metric.  Moreover, it is bounded:
\keep{$\| f\|_1 = \|f\cdot 1\|_1 \le \|f\|_p\|1\|_q =
  \|f\|_p\,\Vol{\Omega}^{1/q}$.  Set~$p=q=2$:
  $\| f\|_1 \le \Vol{\Omega}^{1/2}\|f\|_2$.}
\begin{equation}
  \DJS(\pp_1,\pp_2)
  \le
  \tfrac12(\log 2)\,\lVert\pp_1 - \pp_2\rVert_1 \le \log 2.
\end{equation}
An intuitive interpretation of the JSD is not so straightforward, and will not
be needed here; see for instance~\cite{Endres2003}.

One remark is in order: notice that in~\eqref{eq:fdiv} and~\eqref{eq:fdiv_dot}
there are several divisions by~$\pp_2$, which should rightfully worry the
reader since potentially~$\pp_2$ could vanish at some points, for instance at
the initial time.  However, for any positive time~$\pp_2$ immediately becomes
strictly positive, because of diffusion.  See the discussion in
\citet[p.~161]{Arnold2008} for more careful considerations.

\section{One-dimensional examples}
\label{sec:examples}

To summarize the previous sections: for a velocity field~$\uv(\xv,\t)$ and
diffusion tensor~$\Dbb(\xv,\t)$, we seek solutions to the advection-diffusion
equation~\eqref{eq:FP} with no-flux boundary conditions~\eqref{eq:noflux}.
Then the possible scenarios, in increasing order of complexity, can be
characterized as follows.
\begin{enumerate}
\item If both conditions in~\eqref{eq:incomp_imperm} hold, then the uniform
  density is~$\varphi(\xv)=1/\Vol{\Omega}$.  This is true whether or not~$\uv$
  and~$\Dbb$ are explicitly time-dependent (i.e., autonomous or
  nonautonomous).  In this case the variance evolution
  equation~\eqref{eq:var_evol} is sufficient to directly show convergence to
  the uniform state.
\item If either condition in~\eqref{eq:incomp_imperm} is unsatisfied, then
  there are three subcategories:
  \begin{enumerate}
  \item For~$\uv$ and~$\Dbb$ time-independent (autonomous), any
    initial~$\pp(\xv,\t_0)$ converges to a nonuniform invariant
    density~$\varphi(\xv)$.
  \item For~$\uv$ and~$\Dbb$ time-periodic with period~$\tau$,
    \begin{equation}
      \uv(\xv,\t) = \uv(\xv,\t+\tau),
      \qquad
      \Dbb(\xv,\t) = \Dbb(\xv,\t+\tau),
      \label{eq:uv_Dbb_periodic}
    \end{equation}
    any initial condition~$\pp(\xv,\t_0)$ converges to a periodic limiting
    invariant density~$\varphi(\xv,\t)$,
    with~$\varphi(\xv,\t) = \varphi(\xv,\t + \tau)$.
  \item For~$\uv$ and~$\Dbb$ time-dependent (nonautonomous), any initial
    condition~$\pp(\xv,\t_0)$ converges to a time-dependent limiting invariant
    density~$\varphi(\xv,\t)$.
  \end{enumerate}
  In case 2 the $\ffc$-divergence evolution equation~\eqref{eq:fdiv_dot} can
  be used directly to show convergence to~$\varphi(\xv,\t)$.
\end{enumerate}

Since case 1 is familiar from the traditional view of mixing, we will give
explicit examples for the subcategories of case 2.

\subsection{Example of case 2(a): Convergence to a nonuniform density}

Consider a simple one-dimensional model where the domain~$\Omega = [0,\L]$,
the velocity~$\uv = \Uc\,\xuv$, and~$(\Dbb)_{ij} = \Dc\,\delta_{ij}$,
with~$\Uc$ and~$\Dc$ constants.  Then~\eqref{eq:FP} simplifies to
\begin{equation}
  \pd_\t\pp + \Uc\pd_\xc\pp - \Dc\,\pd_\xc^2\pp = 0,
  \qquad
  0 < \xc < \L
  \label{eq:FP_1D}
\end{equation}
with no-flux boundary conditions
\begin{equation}
  \Uc\pp - \Dc\,\pd_\xc\pp = 0,
  \qquad
  \xc = 0,\L.
  \label{eq:noflux_1D}
\end{equation}
This may be regarded as a simple model of a filter: the flow is
nondivergent and can pass through the membranes at~$\xc=0$ and~$\L$, but
particles cannot cross those membranes.  Since the velocity and diffusivity
are time-independent, \cref{eq:FP_1D} has the invariant density
\begin{equation}
  \varphi(\xc)
  =
  \frac{\Uc}{\Dc}\,\frac{\ee^{\Uc\xc/\Dc}}{\ee^{\Uc\L/\Dc} - 1}\,.
  \label{eq:invdens_1D}
\end{equation}
The flow pushes particle against the boundary at~$\xc=\L$ (for $\Uc>0$),
creating a boundary layer of thickness~$\Dc/\Uc$.

\begin{figure}
  \begin{center}
  \subfigure[]{
    \includegraphics[width=.45\textwidth]{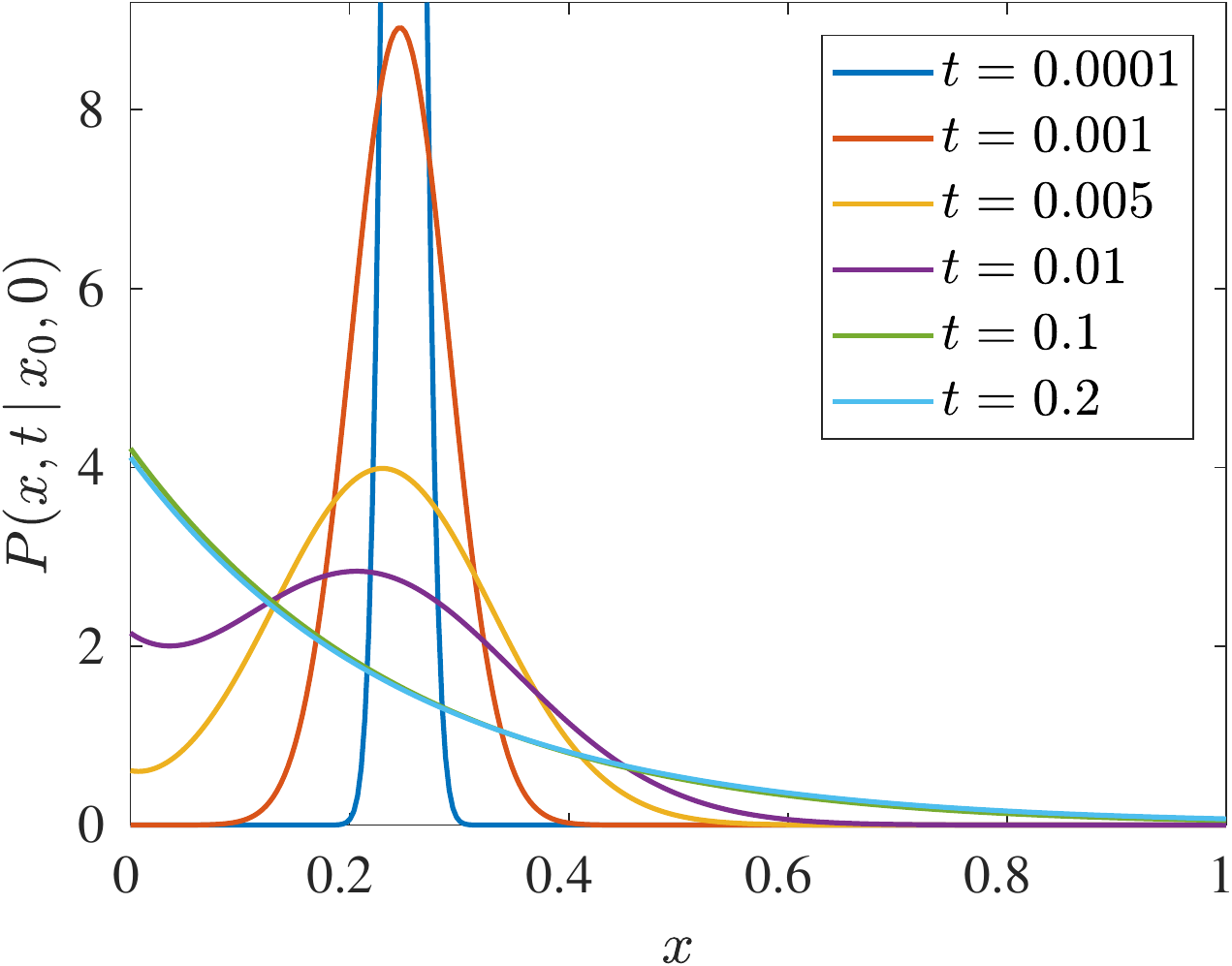}
    \label{fig:intAD_u-4}
  }
  \subfigure[]{
    \includegraphics[width=.45\textwidth]{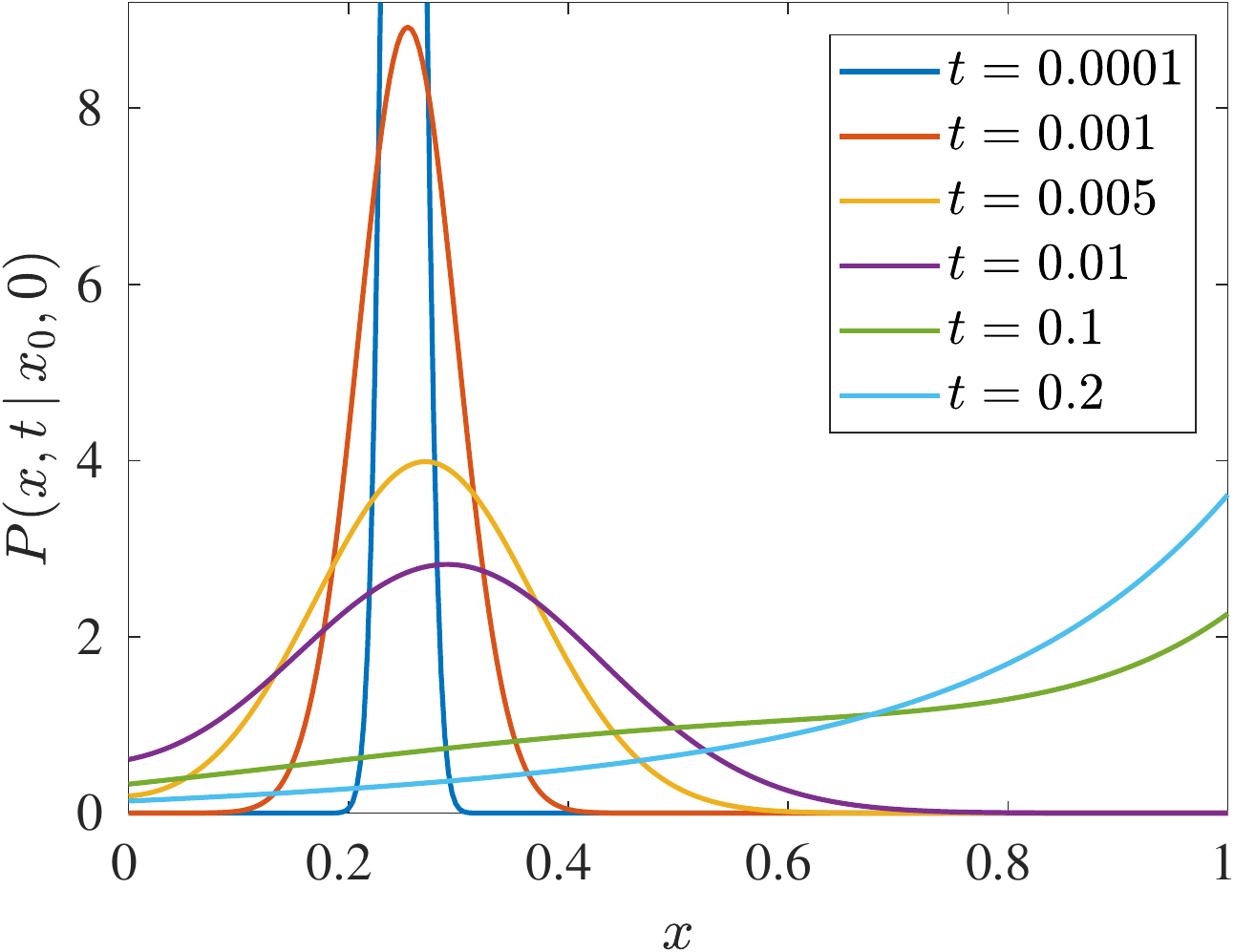}
    \label{fig:intAD_u4}
  }\hspace{.05\textwidth}
  \end{center}
  \caption{The Green's function~\eqref{eq:intADsol} with~$\xc_0=1/4$, plotted
    at different times for~$\Dc=\L=1$ and (a) $\Uc=-4$; (b) $\Uc=4$.  The
    ultimate state is the invariant density~\cref{eq:invdens_1D}.}
  \label{fig:intAD}
\end{figure}

Now we solve the initial value problem for~\cref{eq:FP_1D}.  This is most
generally done in terms of the Green's
function~$\pp=\PP(\xc,\t\,\vert\,\xc_0,\t_0)=\PP(\xc,\t-\t_0\,\vert\,\xc_0,0)$,
which satisfies~\eqref{eq:FP_1D}--\eqref{eq:noflux_1D} with initial
condition~$\PP(\xc,\t_0\,\vert\,\xc_0,\t_0) = \delta(\xc-\xc_0)$.  The
solution is not completely straightforward, since the PDE is not self-adjoint,
but it can be obtained using Laplace transforms as 
\begin{equation}
  \PP(\xc,\t\,\vert\,\xc_0,\t_0)
  =
  \varphi(\xc)
  +
  \frac{\Dc}{\L^3}
  \sum_{n = 1}^\infty
  \frac{\ee^{-\gamma_n(\t-\t_0)}}{2\gamma_n}\,\phi_n^\Uc(\xc)\,\phi_n^{-\Uc}(\xc_0),
  \label{eq:intADsol}
\end{equation}
with
\begin{equation}
  \phi_n^\Uc(\xc)
  =
  \ee^{\Uc\xc/2\Dc}\l\{2\pi n\cos(n\pi x/\L)
  +
  (\lvert\Uc\rvert\L/\Dc)\sin(n\pi\xc/\L)\r\}
\end{equation}
and decay rates
\begin{equation}
  \gamma_n = \Dc\,(\pi n/\L)^2 + \Uc^2/4\Dc.
\end{equation}
The Green's function is plotted in \cref{fig:intAD}.  The relaxation rate to
the invariant density~$\varphi(\xc)$ is given by~$\gamma_1$, which is
considerably enhanced by the constant flow~$\Uc$: the second term~$\Uc^2/4\Dc$
is dominant for~$\Uc\L/\Dc > 2\pi$.  Thus, unlike in `traditional' mixing
problems, a constant velocity can accelerate mixing substantially (though for
a large domain size there could be an initial transient before the
concentration reaches the wall).  This acceleration is due to the flow
squashing the concentration field against the boundary.  Superficially, this
does not sound like mixing, but it is in the sense that it causes the scalar
field~$\pp(\xc,\t)$ to quickly forget its initial condition and converge to
the invariant density~$\varphi(\xc)$.

\subsection{Example of case 2(b): Convergence to a time-periodic density}
\label{sec:example_periodic}

To illustrate convergence to a time-periodic invariant
density~$\varphi(\xc,\t)$, we use the same system
\eqref{eq:FP_1D}--\eqref{eq:noflux_1D} as in the previous example.  We mimic a
time-periodic flow by reversing the direction of~$\uv = \pm\Uc\,\xuv$ at every
half-period~$\tau/2$.  (This could represent the air flow reversing direction
as a mask-wearer inhales and exhales.)  Thus, the density~$\pp(\xc,\t)$ at
time~$\t$ is evolved to time~$\t+\tfrac12\tau$ by
\begin{align}
  \pp(\xc,\t + \tfrac12\tau)
  &=
  \int_0^\L
  \PP_\Uc(\xc,\tfrac12\tau\,\vert\,\xc_0,0)\,\pp(\xc_0,\t)\dint\xc_0
\end{align}
where~$\PP_\Uc$ is the Green's function~\eqref{eq:intADsol}; then, for the
next half-period, we evolve the density with a flow~$\uv=-\Uc\xuv$ to the
left:
\begin{align}
  \pp(\xc,\t + \tau)
  &=
  \int_0^\L
  \PP_{-\Uc}(\xc,\tfrac12\tau\,\vert\,\xc_0',0)\,
  \pp(\xc_0',\t + \tfrac12\tau)\dint\xc_0'
  \nonumber\\
  &=
  \int_0^\L
  \mathcal{P}_\tau(\xc\,\vert\,\xc_0)\,
  \pp(\xc_0,\t)\dint\xc_0
  \label{eq:period-tau}
\end{align}
where the period-$\tau$ kernel is
\begin{equation}
  \mathcal{P}_\tau(\xc\,\vert\,\xc_0)
  \ldef
  \int_0^\L
  \PP_{-\Uc}(\xc,\tfrac12\tau\,\vert\,\xc_0',0)\,
  \PP_{\Uc}(\xc_0',\tfrac12\tau\,\vert\,\xc_0,0)\,
  \dint\xc_0'\,.
  \label{eq:tau_kernel}
\end{equation}

\keep{%
\begin{equation}
  \PP_\Uc(\xc_0',\tfrac12\tau\,\vert\,\xc_0,0)
  =
  \varphi_\Uc(\xc_0')
  +
  \frac{\Dc}{\L^3}
  \sum_{n = 1}^\infty
  \frac{\ee^{-\gamma_n\tau/2}}{2\gamma_n}\,\phi_n^\Uc(\xc_0')\,\phi_n^{-\Uc}(\xc_0),
\end{equation}
\begin{equation}
  \PP_{-\Uc}(\xc,\tfrac12\tau\,\vert\,\xc_0',0)
  =
  \varphi_{-\Uc}(\xc)
  +
  \frac{\Dc}{\L^3}
  \sum_{n = 1}^\infty
  \frac{\ee^{-\gamma_n\tau/2}}{2\gamma_n}\,\phi_n^{-\Uc}(\xc)\,\phi_n^{\Uc}(\xc_0'),
\end{equation}
}

Note that~$\t$ in~\cref{eq:period-tau} is not arbitrary but is aligned with
period boundaries: $\t = \t_k = k\tau$, for integer~$k$.  \Cref{eq:period-tau}
maps the density~$\pp(\xc,\t_k)$ to the beginning of the next period at
time~$\t_{k+1} = (k+1)\tau$.  The invariant density~$\varphi(\xc,\t_k)$ may be
found from
\begin{equation}
  \varphi(\xc,\t_k)
  =
  \int_0^\L
  \mathcal{P}_\tau(\xc\,\vert\,\xc_0)\,
  \varphi(\xc,\t_k)\dint\xc_0
\end{equation}
which is a Fredholm integral equation of the second kind.
Here~$\varphi(\xc,\t_k)$ is the periodic invariant density evaluated at the
start of a period.  Even for this simple time-periodic example it is not
straightforward to compute~$\varphi(\xc,\t_k)$, or the rate of convergence
to~$\varphi(\xc,\t_k)$.

\begin{figure}
  \includegraphics[width=.8\textwidth]{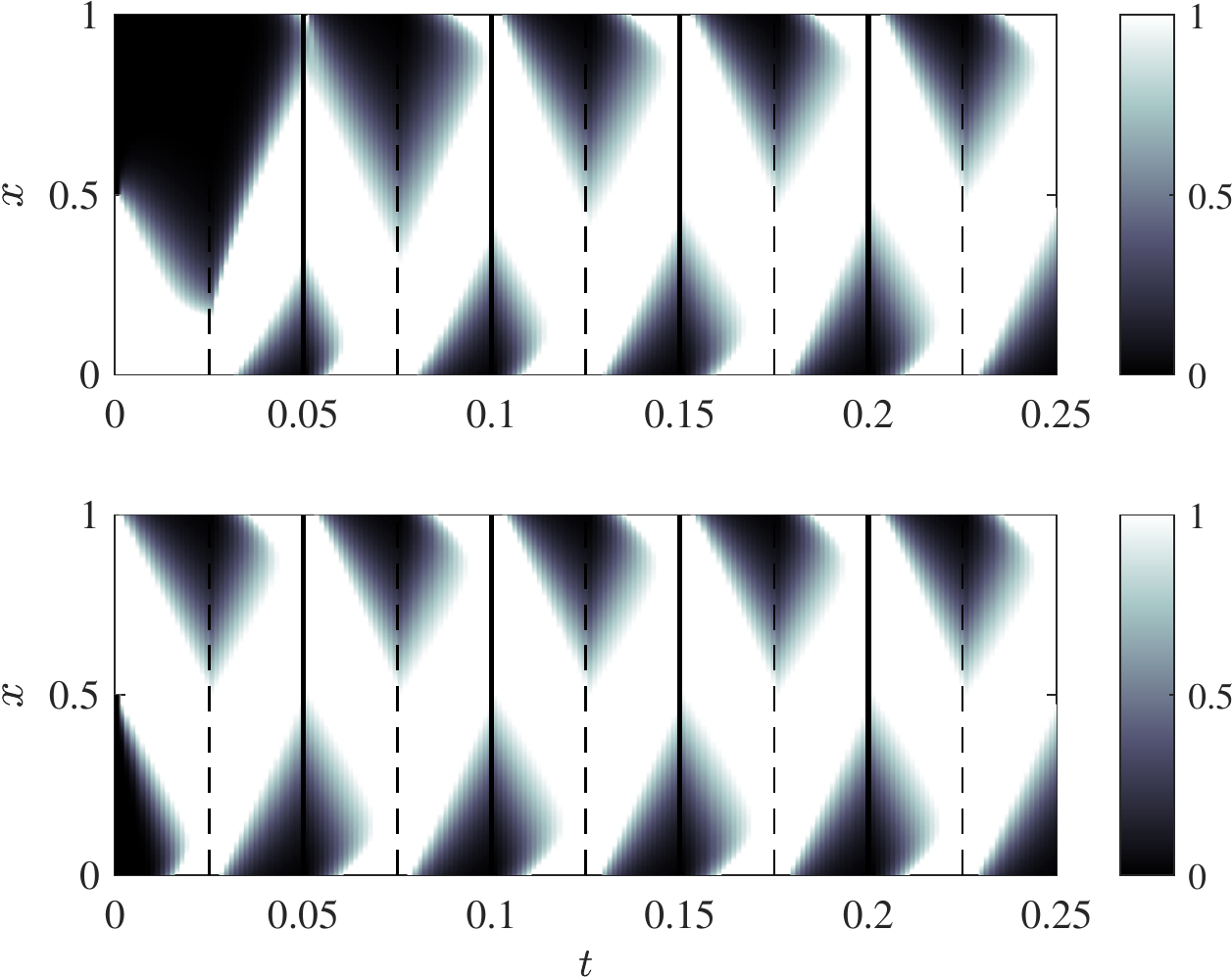}
  \caption{Density~$\pp(\xc,\t)$ for the periodic flow of
    \cref{sec:example_periodic}, for two different initial conditions.  The
    two initial conditions converge to the same periodic
    pattern~$\varphi(\xc,\t)$ after about three periods.  Here the
    period~$\tau=0.05$ and drift~$\Uc=20$, for a domain of width~$\L=1$ and
    with diffusivity~$\Dc=1$.  The vertical lines indicate period boundaries,
    and the dashed lines are half-periods when the flow switches from right to
    left.}
  \label{fig:adnonauto_periodic}
\end{figure}

\begin{figure}
  \includegraphics[width=.6\textwidth]{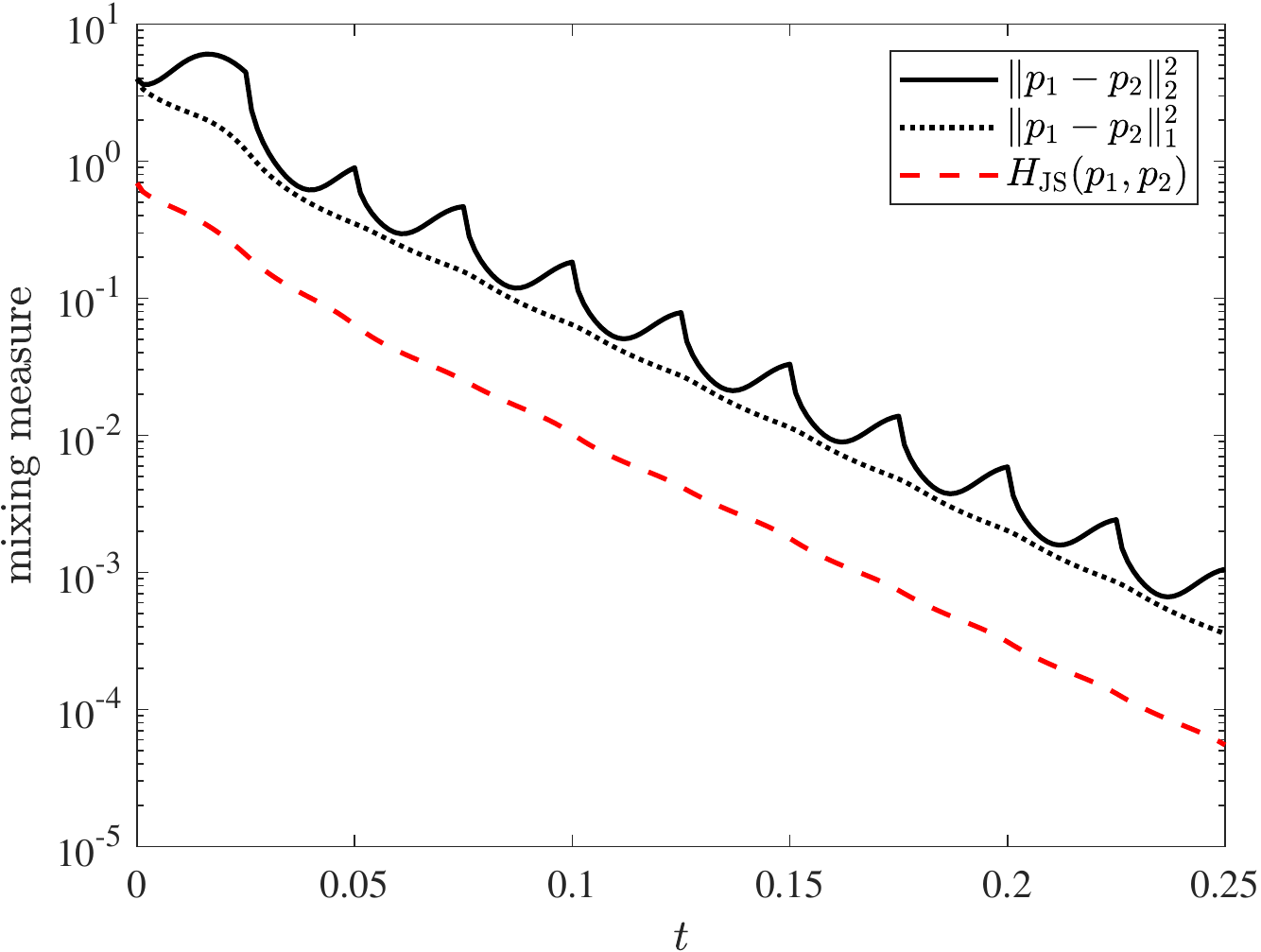}
  \caption{Variance or $L^2$ norm (solid line), squared $L^1$ norm (dotted
    line), and Jensen--Shannon divergence (dashed line) between the two
    solutions in \cref{fig:adnonauto_periodic}.  The variance is nonmonotonic,
    whereas the~$L^1$ norm and~$\DJS$ decreases monotonically.}
  \label{fig:adnonauto_periodic_norms}
\end{figure}

In \cref{fig:adnonauto_periodic} we show a numerical solution of
\cref{eq:period-tau}, for two different initial conditions: the first
($\pp_1(\xc,0)$) has particles initially concentrated on the right side of the
interval, and the second ($\pp_2(\xc,0)$) on the left.  The two solutions
rapidly converge to each other after about three periods.  The ultimate
state~$\varphi(\xc,\t)$ may be considered `mixed' even if it is not uniform.
In \cref{fig:adnonauto_periodic_norms} we compare the time evolution of
variance~$\int_\Omega\lvert\pp_1 - \pp_2\rvert^2\dV$ to the Jensen--Shannon
divergence~\eqref{eq:JSD}.  The variance is not at all monotonic: it
oscillates about a decreasing trend.  The JSD is nice and monotonic, which
makes it much easier to assign a numerical value to the decay rate.

\subsection{Example of case 2(c): Convergence to an aperiodic density}
\label{sec:example_random}

A simple way to produce an example that is neither steady nor time-periodic is
to add some randomness~\cite{Pierrehumbert1994}.  Recall that in the periodic
example of \cref{sec:example_periodic} we imposed a flow~$\Uc\xuv$ to the
right for a time~$\tfrac12\tau$, followed by a flow~$-\Uc\xuv$ to the left for
a time~$\tfrac12\tau$, to obtain a period-$\tau$ map.  One simple way to
randomize this process is to select for every time interval $[\t_k,\t_k+\tau)$
a uniform independent random number~$\alpha_k \in [0,1]$, and impose a flow to
the right for a time~$\alpha_k\tau$, followed by a flow to the left for a
time~$(1-\alpha_k)\tau$.  The kernel \cref{eq:tau_kernel} is then replaced by
\begin{equation}
  \mathcal{P}_{\tau,\alpha_k}(\xc\,\vert\,\xc_0)
  \ldef
  \int_0^\L
  \PP_{-\Uc}(\xc,(1-\alpha_k)\tau\,\vert\,\xc_0',0)\,
  \PP_{\Uc}(\xc_0',\alpha_k\tau\,\vert\,\xc_0,0)\,
  \dint\xc_0'
\end{equation}
and the map from time~$\t_k$ to~$\t_{k+1} = \t_k + \tau$ is
\begin{equation}
  \pp(\xc,\t_k + \tau)
  =
  \int_0^\L
  \mathcal{P}_{\tau,\alpha_k}(\xc\,\vert\,\xc_0)\,
  \pp(\xc_0,\t_k)\dint\xc_0\,.
  \label{eq:Pc_random}
\end{equation}

\begin{figure}
  \includegraphics[width=.8\textwidth]{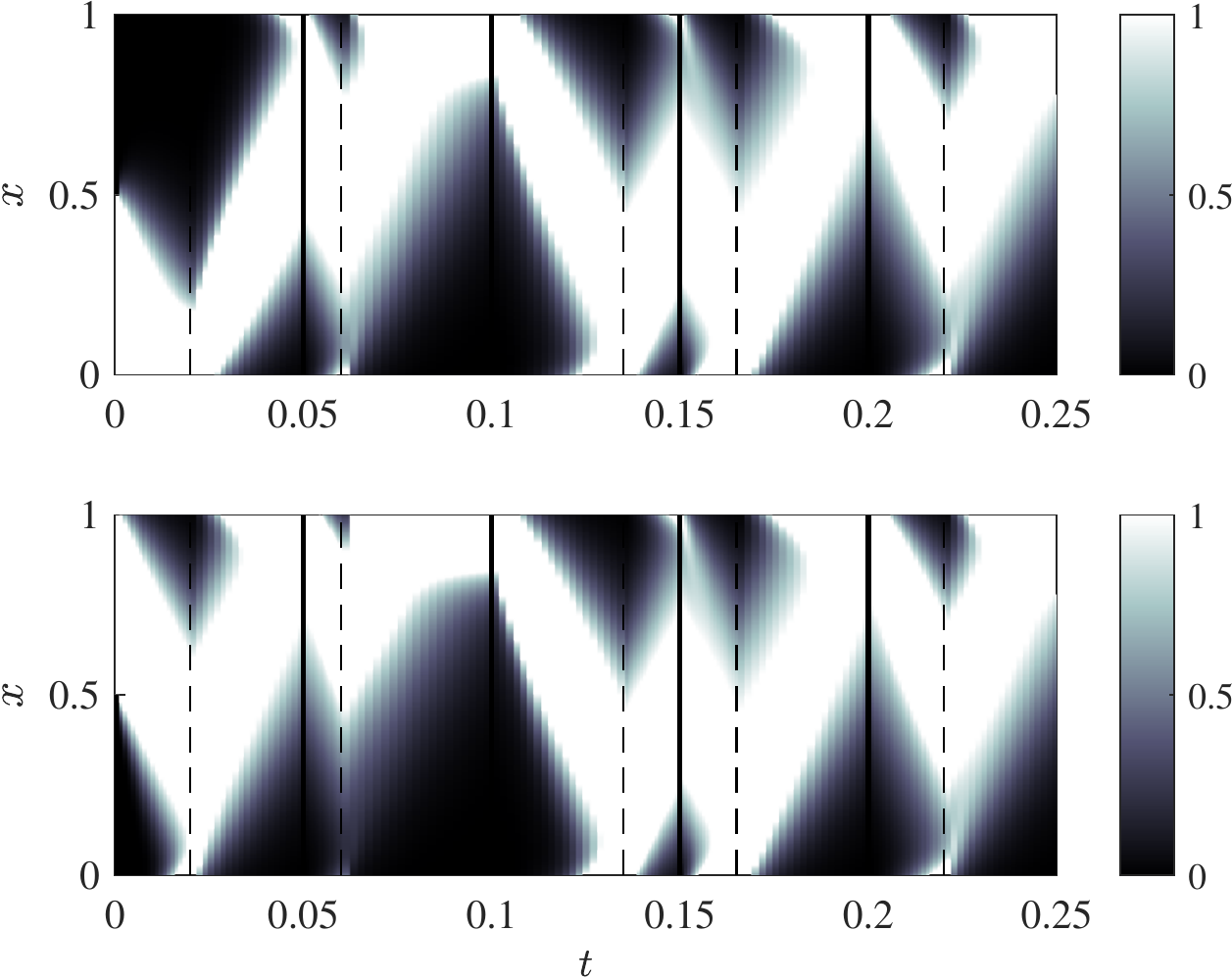}
  \caption{Density~$\pp(\xc,\t)$ for the random flow of
    \cref{sec:example_random}, for two different initial conditions.  The two
    initial conditions rapidly converge to the same random
    pattern~$\varphi(\xc,\t)$.  The vertical lines indicate period
    boundaries~$\t_k$, and the dashed lines are the random
    times~$\t_k + \alpha_k\tau$ when the flow switches from right to left.
    Parameter values are as in \cref{fig:adnonauto_periodic}.}
  \label{fig:adnonauto_random}
\end{figure}

\begin{figure}
  \includegraphics[width=.6\textwidth]{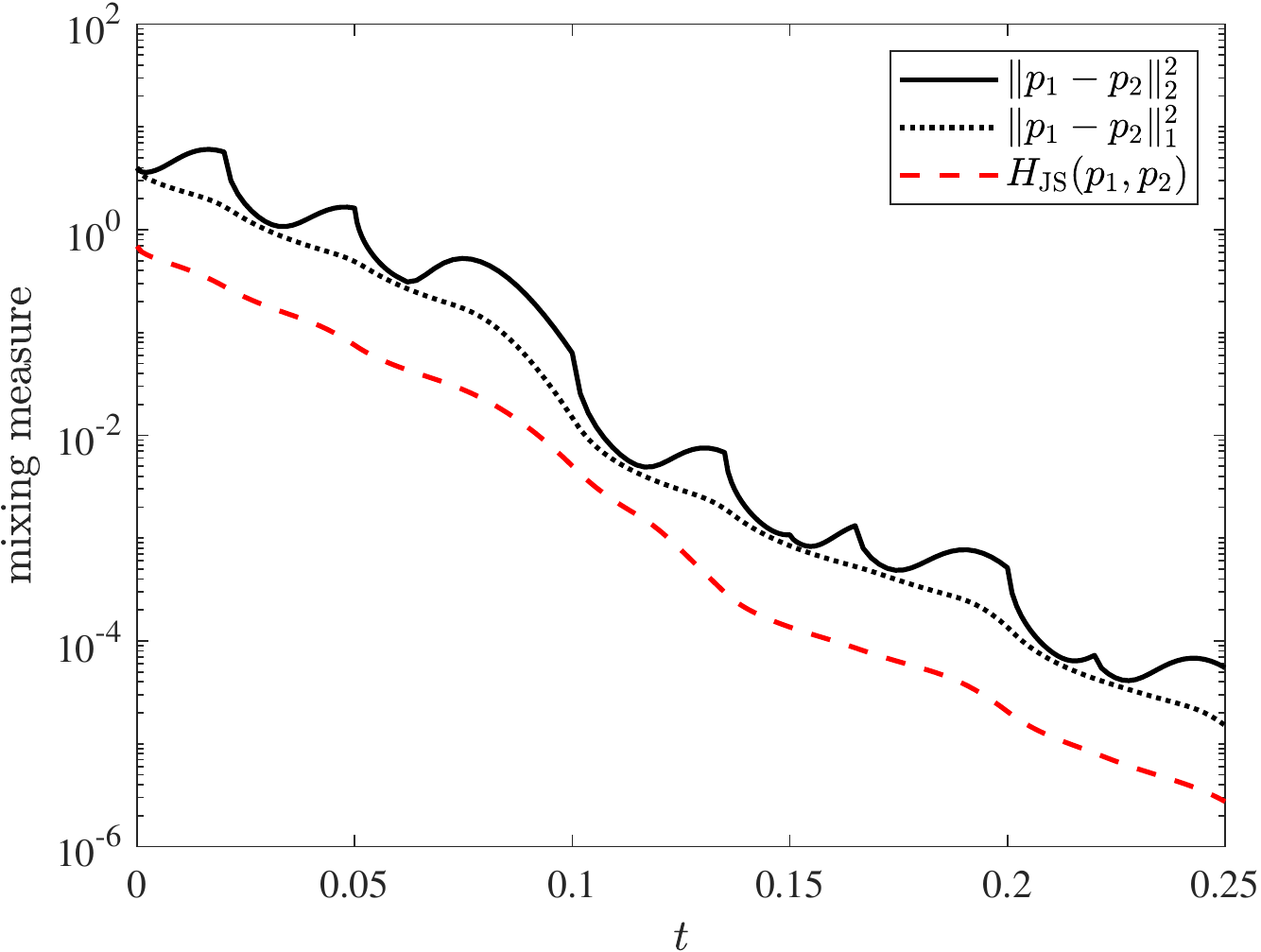}
  \caption{Variance or $L^2$ norm (solid line), squared $L^1$ norm (dotted
    line), and Jensen--Shannon divergence (dashed line) between the two
    solutions in \cref{fig:adnonauto_random}.  The variance is nonmonotonic,
    whereas the~$L^1$ norm and~$\DJS$ decreases monotonically.}
  \label{fig:adnonauto_random_norms}
\end{figure}

In \cref{fig:adnonauto_random} we show a numerical solution of
\cref{eq:Pc_random}, for two different initial conditions, which rapidly
converge to each other after about three periods.  The ultimate
state~$\varphi(\xc,\t)$ is `mixed' even though it is neither uniform nor
periodic.  In \cref{fig:adnonauto_random_norms} we compare the time evolution
of variance~$\int_\Omega\lvert\pp_1 - \pp_2\rvert^2\dV$ to the Jensen--Shannon
divergence~\eqref{eq:JSD}.  Much like the periodic case, the variance is not
at all monotonic, whereas the JSD relentlessly decreases towards zero.

\keep{%
\begin{equation}
  \PP_\Uc(\xc_0',\alpha_k\tau\,\vert\,\xc_0,0)
  =
  \varphi_\Uc(\xc_0')
  +
  \frac{\Dc}{\L^3}
  \sum_{n = 1}^\infty
  \frac{\ee^{-\gamma_n\alpha_k\tau}}
  {2\gamma_n}\,\phi_n^\Uc(\xc_0')\,\phi_n^{-\Uc}(\xc_0),
\end{equation}
\begin{equation}
  \PP_{-\Uc}(\xc,(1 - \alpha_k)\tau\,\vert\,\xc_0',0)
  =
  \varphi_{-\Uc}(\xc)
  +
  \frac{\Dc}{\L^3}
  \sum_{n = 1}^\infty
  \frac{\ee^{-\gamma_n(1 - \alpha_k)\tau}}
  {2\gamma_n}\,\phi_n^{-\Uc}(\xc)\,\phi_n^{\Uc}(\xc_0'),
\end{equation}
}

\section{Sources and sinks}
\label{sec:sources}

\subsection{Varying the number of particles}
\label{sec:varying_particles}

So far the number of particles was fixed.  Now consider the particle
density~$\nn(\xv,\t) \ge 0$ (also sometimes called particle number or number
density), which obeys the equation
\begin{equation}
  \pd_\t\nn + \div\Fv(\nn) = \Q(\xv,\t;\nn),
  \qquad
  \xv \in \Omega,
  \label{eq:nn}
\end{equation}
with the particle flux~$\Fv(\nn) = \uv\nn - \Dbb\cdot\grad\nn$ defined as
in~\eqref{eq:Fv}.  The particle density differs from the probability
density~$\pp(\xv,\t)$ in that the number of particles
\begin{equation}
  \N(\t) = \int_\Omega \nn(\xv,\t)\dV
  \label{eq:N}
\end{equation}
is not necessarily~$1$, and can change with time.  The number of
particles~$\N(\t)$ is not in general an integer.  This can either be
interpreted as a small error when~$\N$ is very large, or~$\nn/\N$ can be
interpreted as a probability.

The source-sink function~$\Q(\xv,\t;\nn)$ is not completely arbitrary: it must
preserve the positivity of~$\nn$.  There is an asymmetry between adding and
removing particles: we can always add particles, but we can only remove
particles if there are particles present.  One common form of~$\Q$ that
naturally enforces this is
\begin{equation}
  \Q(\xv,\t;\nn) = \S(\xv,\t) - \K(\xv,\t)\,\nn,
  \qquad
  \S \ge 0,\ \K \ge 0
  \label{eq:Q}
\end{equation}
for given nonnegative functions~$\S$ and~$\K$.  The source~$\S$ creates
particles indiscriminately, but the sink~$-\K\nn$ vanishes
as~$\nn \rightarrow 0$.  Of course, more general forms than~\eqref{eq:Q} are
possible.

The same considerations for the interior source-sink apply to the flux at the
boundary: we should not remove particles if there are none present.  Thus, we
write for the boundary flux
\begin{equation}
  \q(\xv,\t;\nn)
  =
  -\Fv(\nn)\cdot\nuv = \s(\xv,\t) - \k(\xv,\t)\,\nn,
  \qquad
  \s \ge 0,\ \k \ge 0,
  \qquad
  \xv \in \pd\Omega
  \label{eq:q}
\end{equation}
for given nonnegative boundary functions~$\s$ and~$\k$.  The minus sign in
front of~$\Fv\cdot\nuv$ in~\eqref{eq:q} is because~$\nuv$ is an outward
normal, so~$\Fv(\nn)\cdot\nuv > 0$ corresponds to particles leaving the
domain~$\Omega$.

Using~\cref{eq:N,eq:nn} and the definition of~$\q$ in~\eqref{eq:q}, we see
that the time evolution of~$\N$ satisfies
\begin{equation}
  \dot\N
  =
  \int_\Omega \Q\dV + \int_{\pd\Omega} \q\dS
\end{equation}
where the first term is the `bulk' source of particles, and the second is the
flux of particles across the boundary of~$\Omega$.

\keep{
We can interpret~$\nn$ as a probability by letting~$\pp = \nn/\N$, which then
obeys
\begin{equation}
  \pd_\t\pp + \div\Fv(\pp)
  = (\Q - \dot\N\,\pp)/\N
  \label{eq:FP_q_over_NN}
\end{equation}
It is easy to check that~\eqref{eq:FP_q_over_NN}
preserves~$\int_\Omega\pp\dV=1$.  The probability density~$\pp$ gives the
relative fraction of particles in a small volume, but the total number of
particles~$\N(\t)$ must be tracked separately.
}

\subsection{Convergence to asymptotic state}

In \cref{sec:comp_or_perm} we showed that, for one particle (or equivalently a
fixed number of noninteracting particles) we can use the \fdiv\ to prove that
any two initial conditions will converge to the same ultimate
state~$\varphi(\xv,\t)$.  The ultimate state may be nonuniform and
time-dependent, but what characterizes it is that it is independent of the
initial condition: it is an asymptotic state.

Having now allowed for sources and sinks in \cref{sec:varying_particles}, we
can ask about defining~$\varphi(\xv,\t)$ in that case.  After all, adding and
removing particles should not prevent two arbitrary initial conditions from
converging to each other, as long as they are subjected to the same sources
and sinks.

We define the difference~$\theta = \nn_1 - \nn_2$ between any two solutions of
\cref{eq:nn}.  The squared-integral of~$\theta$ obeys an equation analogous to
the variance evolution~\cref{eq:var_evol}:
\begin{multline}
  \frac{\d}{\dt}\int_\Omega\theta^2 \dV
  =
  \int_\Omega
  \uv\cdot\grad\theta^2\dV
  -
  2\int_\Omega\grad\theta\cdot\Dbb\cdot\grad\theta\dV
  \\
  -
  2\int_\Omega\K\,\theta^2\dV
  -
  2\int_{\pd\Omega}\k\,\theta^2\dV.
  \label{eq:var_evol_nn}
\end{multline}
The source~$\S$ does not enter the equation; the last two terms are new but
they are nonpositive, so they promote convergence to an equilibrium.  However,
the same sign-indefinite term involving the integral
of~$\uv\cdot\grad\theta^2$ appears on the right.  This term does go away under
the nondivergence and impermeability assumptions~\eqref{eq:incomp_imperm}, in
which case~\eqref{eq:var_evol_nn} is enough to conclude convergence to an
ultimate state~$\varphi(\xv,\t)$, independent of initial condition.

However, in the divergent or permeable case, we have the same problem as
before: the presence of a sign-indefinite term prevents us from guaranteeing
convergence.  A generalization of the~\fdiv~\eqref{eq:fdiv} is needed, with a
time evolution that allows us to conclude convergence.  We tentatively define
\begin{equation}
  \Df[\nn_1,\nn_2] =
  \int_{\Omega}
  \nn_2\,\ffc(\nn_1/\nn_2)\dV.
  \label{eq:fdiv_nn}
\end{equation}
This is not strictly speaking an \fdiv, since~$\nn_1$ and~$\nn_2$ are not
normalized probability densities.  The proof from \cref{eq:Df_pos} that~$\Df$
is positive now reads
\begin{equation*}
  \Df[\nn_1,\nn_2]
  =
  \N_2
  \int_{\Omega}
  \ffc(\nn_1/\nn_2)\,\pp_2\dV
  \ge
  \N_2
  \ffc\l(\int_\Omega (\nn_1/\nn_2)\,\pp_2 \dV\r)
  =
  \N_2\ffc(\N_1/\N_2)
\end{equation*}
where~$\N_i = \int_\Omega\nn_i\dV$, and~$\pp_i=\nn_i/\N_i$ are normalized
probability densities.  Hence, to guarantee~$\Df[\nn_1,\nn_2] \ge 0$ we must
add the additional requirement that~$\ffc \ge 0$, which is satisfied by the
Jensen--Shannon choice~\eqref{eq:ffc_JS} for~$\ffc$.  (To satisfy~$\ffc\ge0$,
the Kullback--Leibler choice~\eqref{eq:ffc_KL} can simply be modified to
read~$\ffc(\uu) = \uu\log\uu - \uu + 1$, in which case~$\Df$ is sometimes
called the \emph{physical relative entropy}.)  With this additional constraint
on~$\ffc$, we have~$\Df[\nn_1,\nn_2]=0$ if and only if~$\nn_1\equiv\nn_2$.

With the same approach as in \cref{apx:fdiv_dot} we can show that the time
evolution of~$\Df[\nn_1,\nn_2]$ is given by
\begin{multline}
  \skew{-3}\dot\Df[\nn_1,\nn_2]
  =
  -
  \int_{\Omega}
  \nn_2\,\ffc''(\nn_1/\nn_2)\,
  \grad(\nn_1/\nn_2)\cdot\Dbb\cdot\grad(\nn_1/\nn_2)
  \dV
  \\
  - \int_\Omega \l(
  \K\nn_2\ffc(\nn_1/\nn_2)
  +
  \S\,\hapf(\nn_1/\nn_2)
  \r)\dV
  \\
  - \int_{\pd\Omega} \l(
  \k\nn_2\ffc(\nn_1/\nn_2)
  +
  \s\,\hapf(\nn_1/\nn_2)
  \r)\dS
  \le 0,
  \label{eq:dot_Df_nn}
\end{multline}
where
\begin{equation}
  \hapf(\uu) \ldef (\uu - 1)\ffc'(\uu) - \ffc(\uu) \ge 0,
  \qquad
  \hapf(1) = 0.
  \label{eq:hapf}
\end{equation}
The inequality in~\eqref{eq:dot_Df_nn} follows from the positivity of~$\nn_i$,
the strict convexity of~$\ffc$ ($\ffc''>0$), the positive-definiteness
of~$\Dbb$, the nonnegativity of~$\ffc$, $\K$, $\S$, $\k$, $\s$, and the
inequality in~\eqref{eq:hapf}.  (The latter is easy to prove: a differentiable
convex function satisfies~$\ffc(x) \ge \ffc(y) + (x-y)f'(y)$ for all~$x$, $y$,
since its graph is above all its tangents; set~$x=1$ and~$y=u$ and
use~$\ffc(1)=0$.)  The right-hand side of~\eqref{eq:dot_Df_nn} vanishes if and
only if~$\nn_1=\nn_2$.

\keep{Can we bound $\int_\Omega\hapf(\nn_1/\nn_2) \ge C\Vol{\Omega}\Df?$}

\section{Total variation distance and~\texorpdfstring{$L^1$}{L1} norm}
\label{sec:TV}

As an alternative to the \fdiv, another measure of convergence of two
densities is the \emph{total variation distance} (or \emph{variational
  distance}), which is equivalent to~$\tfrac12\lVert\pp_1 - \pp_2\rVert_1$
\cite{CoverThomas}, where~$\lVert\cdot\rVert_1$ is the~$L^1$ norm on~$\Omega$.
Compare the evolution of \hbox{$\lVert\pp_1 - \pp_2\rVert_1^2$} to the
concentration variance~$\lVert\pp_1 - \pp_2\rVert_2^2$ in
\cref{fig:adnonauto_periodic_norms,fig:adnonauto_random_norms}.  Notice that
the~$L^1$ norm, much like~$\Df$, decays monotonically, exhibiting none of the
troublesome oscillations of the~$L^2$ norm (variance).  In this section we
will show that the~$L^1$ norm does indeed always decrease monotonically, so
that it is a more reliable measure of mixing than the~$L^2$ norm for
nonuniform mixing.

We shall prove this for two general number densities~$\nn_1$ and~$\nn_2$
obeying \cref{eq:nn} with the source-sink~\eqref{eq:Q}, and with boundary
conditions~\eqref{eq:q}.  Let~$\theta = \nn_1 - \nn_2$, which
satisfies~$\pd_\t\theta = -\div\Fv(\theta) - \K\theta$
and~$\Fv(\theta)\cdot\nuv = \k\,\theta$ on~$\pd\Omega$.  For any
function~$G(\theta)$, we have
\begin{multline}
  \frac{\d}{\dt}\int_\Omega G(\theta)\dV
  =
  \int_\Omega G''(\theta)\,\theta\uv\cdot\grad\theta\dV
  -
  \int_\Omega G''(\theta)\,\grad\theta\cdot\Dbb\cdot\grad\theta\dV \\
  -
  \int_\Omega G'(\theta)\,\K\theta\dV
  -
  \int_{\pd\Omega} G'(\theta)\,\k\theta\dS
  \label{eq:dt_intG}
\end{multline}
which is a generalization of \cref{eq:var_evol_nn}.  Now
let~$G(\theta) = \lvert\theta\rvert$, so that~$G'(\theta) = \sgn(\theta)$
and~$G''(\theta) = 2\delta(\theta)$.  With that choice, \cref{eq:dt_intG}
becomes
\begin{equation*}
  \frac{\d}{\dt}\lVert\theta\rVert_1
  =
  2\int_\Omega \delta(\theta)\,\theta\uv\cdot\grad\theta\dV
  -
  2\int_\Omega \delta(\theta)\,\grad\theta\cdot\Dbb\cdot\grad\theta\dV
  -
  \int_\Omega \K\lvert\theta\rvert\dV
  -
  \int_{\pd\Omega} \k\lvert\theta\rvert\dS.
\end{equation*}
The first term on the right vanishes since the delta function
forces~$\theta=0$; for the second term, we can turn the volume integral into a
surface integral \cite[Theorem 6.1.5]{HormanderI}:
\begin{align}
  \frac{\d}{\dt}\lVert\theta\rVert_1
  &=
  -2\int_{\{\theta=0\}} \grad\theta\cdot\Dbb\cdot\grad\theta
  \,\frac{\dS}{\lvert\grad\theta\rvert}
  -
  \int_\Omega \K\lvert\theta\rvert\dV
  -
  \int_{\pd\Omega} \k\lvert\theta\rvert\dS
  \le 0,
  \label{eq:L1_evol}
\end{align}
where the first integral is over the zero level set of~$\theta(\cdot,\t)$.  We
conclude that the total variation distance~$\tfrac12\lVert\nn_1-\nn_2\rVert_1$
does indeed decrease monotonically, as was apparent from the earlier numerical
simulations.  (The level-set integral in \cref{eq:L1_evol} appears in
approaches based on tracer coordinates~\cite{Nakamura1996}.)

The `proof' presented here relies on the apparently strong assumption
that~$\lvert\grad\theta\rvert \ne 0$ on the zero level set of~$\theta$.
However, the uniform ellipticity bound~\eqref{eq:uniform_elliptic} implies
that~$-\grad\theta\cdot\Dbb\cdot\grad\theta/\lvert\grad\theta\rvert \le
-\sigma\lvert\grad\theta\rvert$, so singular points limit nicely to zero in
the integrand.  In \cref{apx:Lq_decay} we show that the~$L^1$ norm is the
only~$L^\qn$ norm that decays monotonically in the nonuniform mixing case.

One possible advantage \cref{eq:L1_evol} has over the corresponding
equation~\eqref{eq:dot_Df_nn} for the \fdiv\ is that it shows convergence even
when the source~$\S(\xv,\t)$ is negative, since the source has dropped out
of~\eqref{eq:L1_evol} completely.  However~\cref{eq:dot_Df_nn} suggests that a
positive source can actually improve the rate of convergence.  Another
weakness of \cref{eq:L1_evol} compared to~\eqref{eq:dot_Df_nn} is that the its
right-hand side is difficult to compute: it requires tracking of the zero
level set, which is a challenging problem in practice because of resolution and
changes in topology.  By comparison, the right-hand side
of~\cref{eq:dot_Df_nn} is readily computed and regions of large entropy
production can be identified from the magnitude of the integrands.

\section{Discussion}
\label{sec:disc}

The traditional view of mixing in nondivergent flow is that a stirred passive
scalar will ultimately be homogenized to a uniform concentration.  As we
discussed, this requires both nondivergence of the velocity field and
no-penetration boundary conditions.  If either condition is violated, the
ultimate state of the mixing process is no longer uniform, and may in fact be
time-dependent for nonautonomous systems, where~$\uv$ or~$\Dbb$ are explicit
functions of time.  We refer to these systems as nonuniform mixing, because
the passive scalar may be mixed even though its concentration is not uniform.
Such nonuniform situations will arise in the presence of filters, which are
membranes that permits the passage of fluid but not of particles (passive
scalar).

Using the standard concentration variance as a proxy for mixing is less useful
for nonuniform mixing, since the variance is not necessarily a
monotonically-decreasing function of time.  Of course, variance will
eventually decrease to zero even in nonuniform mixing (as long as it it
defined appropriately), but the excursions it undertakes can make it hard to
ascribe a rate of mixing to the system (see
\cref{fig:adnonauto_periodic_norms,fig:adnonauto_random_norms}).  Instead of
concentration variance, a more reliable proxy for mixing is the \emph{\fdiv},
which is related to relative entropy.  Instead of relying an initial condition
to become uniform, we define the rate of mixing in terms of the rate at which
two arbitrary densities~$\pp_1(\xv,\t)$ and~$\pp_2(\xv,\t)$ approach each
other.  They will eventually both converge to an ultimate
density~$\varphi(\xv,\t)$, which is independent of the initial condition.  The
\fdiv\ picture is easily adapted to cases with sources and sinks.

The connection between the \fdiv\ and mix-norms \cite{Mathew2003, Mathew2005,
  Thiffeault2012} is not completely clear.  Mix-norms are used as a diagnostic
for mixing, and are not guaranteed to decay monotonically for the types of
examples presented here.  Their behavior is thus probably more closely related
to that of concentration variance than to \fdiv, though they have the
advantage that they decay even when the diffusivity is set to zero, which
renders them more useful for optimization \cite{Mathew2007,
  Lin2011b,Foures2014, Vermach2018, Marcotte2018b}.  Perhaps there is a hybrid
approach that could marry the advantages of both.

Finally, note that nonuniform mixing suggests a different type of mixing
optimization problem, where the goal is to decrease spatial or temporal
variations of~$\varphi(\xv,\t)$ itself rather than the rate of approach
to~$\varphi(\xv,\t)$.  This was investigated previously for source-sink
systems \cite{Thiffeault2004, DoeringThiffeault2006, Shaw2007,
  Thiffeault2008}, but it could be effected in any problem involving
nonuniform mixing.  For example, a flow could be designed to minimize the
concentration of viruses near a filter, to mitigate the effect of
inevitable imperfections in the membrane.

\begin{acknowledgments}
  The author thanks Yu Feng, Albion Lawrence, Noboru Nakamura, Bryan Oakley,
  Greg Pavliotis, Jim Thomas, Jeffrey Weiss, Bill Young, and an anonymous
  referee for insightful comments and discussions.  Some of this research was
  completed at the Aspen Center for Physics, which is supported by National
  Science Foundation grant PHY-1607611; travel there was supported by the
  Brandeis University Provost's Research Grant ``Nonequilibrium Statistical
  Mechanics of the Ocean and Atmosphere.''
\end{acknowledgments}

\bibliography{journals_abbrev,articles}

\begin{thebibliography}{51}%
\makeatletter
\providecommand \@ifxundefined [1]{%
 \@ifx{#1\undefined}
}%
\providecommand \@ifnum [1]{%
 \ifnum #1\expandafter \@firstoftwo
 \else \expandafter \@secondoftwo
 \fi
}%
\providecommand \@ifx [1]{%
 \ifx #1\expandafter \@firstoftwo
 \else \expandafter \@secondoftwo
 \fi
}%
\providecommand \natexlab [1]{#1}%
\providecommand \enquote  [1]{``#1''}%
\providecommand \bibnamefont  [1]{#1}%
\providecommand \bibfnamefont [1]{#1}%
\providecommand \citenamefont [1]{#1}%
\providecommand \href@noop [0]{\@secondoftwo}%
\providecommand \href [0]{\begingroup \@sanitize@url \@href}%
\providecommand \@href[1]{\@@startlink{#1}\@@href}%
\providecommand \@@href[1]{\endgroup#1\@@endlink}%
\providecommand \@sanitize@url [0]{\catcode `\\12\catcode `\$12\catcode
  `\&12\catcode `\#12\catcode `\^12\catcode `\_12\catcode `\%12\relax}%
\providecommand \@@startlink[1]{}%
\providecommand \@@endlink[0]{}%
\providecommand \url  [0]{\begingroup\@sanitize@url \@url }%
\providecommand \@url [1]{\endgroup\@href {#1}{\urlprefix }}%
\providecommand \urlprefix  [0]{URL }%
\providecommand \Eprint [0]{\href }%
\providecommand \doibase [0]{https://doi.org/}%
\providecommand \selectlanguage [0]{\@gobble}%
\providecommand \bibinfo  [0]{\@secondoftwo}%
\providecommand \bibfield  [0]{\@secondoftwo}%
\providecommand \translation [1]{[#1]}%
\providecommand \BibitemOpen [0]{}%
\providecommand \bibitemStop [0]{}%
\providecommand \bibitemNoStop [0]{.\EOS\space}%
\providecommand \EOS [0]{\spacefactor3000\relax}%
\providecommand \BibitemShut  [1]{\csname bibitem#1\endcsname}%
\let\auto@bib@innerbib\@empty
\bibitem [{\citenamefont {Thiffeault}(2008)}]{ThiffeaultAosta2004}%
  \BibitemOpen
  \bibfield  {author} {\bibinfo {author} {\bibfnamefont {J.-L.}\ \bibnamefont
  {Thiffeault}},\ }\bibfield  {title} {\bibinfo {title} {Scalar decay in
  chaotic mixing},\ }in\ \href {https://doi.org/10.1007/978-3-540-75215-8_1}
  {\emph {\bibinfo {booktitle} {Transport and Mixing in Geophysical Flows}}},\
  \bibinfo {series} {Lecture Notes in Physics}, Vol.\ \bibinfo {volume} {744},\
  \bibinfo {editor} {edited by\ \bibinfo {editor} {\bibfnamefont {J.~B.}\
  \bibnamefont {Weiss}}\ and\ \bibinfo {editor} {\bibfnamefont
  {A.}~\bibnamefont {Provenzale}}}\ (\bibinfo  {publisher} {Springer},\
  \bibinfo {address} {Berlin},\ \bibinfo {year} {2008})\ pp.\ \bibinfo {pages}
  {3--35},\ \Eprint {https://arxiv.org/abs/arXiv:nlin/0502011}
  {arXiv:nlin/0502011} \BibitemShut {NoStop}%
\bibitem [{\citenamefont {Aref}\ \emph {et~al.}(2017)\citenamefont {Aref},
  \citenamefont {Blake}, \citenamefont {Budi{\v s}i{\'c}}, \citenamefont
  {Cardoso}, \citenamefont {Cartwright}, \citenamefont {Clercx}, \citenamefont
  {El~Omari}, \citenamefont {Feudel}, \citenamefont {Golestanian},
  \citenamefont {Gouillart}, \citenamefont {{van Heijst}}, \citenamefont
  {Krasnopolskaya}, \citenamefont {Le~Guer}, \citenamefont {MacKay},
  \citenamefont {Meleshko}, \citenamefont {Metcalfe}, \citenamefont
  {Mezi{\'c}}, \citenamefont {{de Moura}}, \citenamefont {Piro}, \citenamefont
  {Speetjens}, \citenamefont {Sturman}, \citenamefont {Thiffeault},\ and\
  \citenamefont {Tuval}}]{Aref2017}%
  \BibitemOpen
  \bibfield  {author} {\bibinfo {author} {\bibfnamefont {H.}~\bibnamefont
  {Aref}}, \bibinfo {author} {\bibfnamefont {J.~R.}\ \bibnamefont {Blake}},
  \bibinfo {author} {\bibfnamefont {M.}~\bibnamefont {Budi{\v s}i{\'c}}},
  \bibinfo {author} {\bibfnamefont {S.~S.}\ \bibnamefont {Cardoso}}, \bibinfo
  {author} {\bibfnamefont {J.~H.}\ \bibnamefont {Cartwright}}, \bibinfo
  {author} {\bibfnamefont {H.~J.}\ \bibnamefont {Clercx}}, \bibinfo {author}
  {\bibfnamefont {K.}~\bibnamefont {El~Omari}}, \bibinfo {author}
  {\bibfnamefont {U.}~\bibnamefont {Feudel}}, \bibinfo {author} {\bibfnamefont
  {R.}~\bibnamefont {Golestanian}}, \bibinfo {author} {\bibfnamefont
  {E.}~\bibnamefont {Gouillart}}, \bibinfo {author} {\bibfnamefont {G.~F.}\
  \bibnamefont {{van Heijst}}}, \bibinfo {author} {\bibfnamefont {T.~S.}\
  \bibnamefont {Krasnopolskaya}}, \bibinfo {author} {\bibfnamefont
  {Y.}~\bibnamefont {Le~Guer}}, \bibinfo {author} {\bibfnamefont {R.~S.}\
  \bibnamefont {MacKay}}, \bibinfo {author} {\bibfnamefont {V.~V.}\
  \bibnamefont {Meleshko}}, \bibinfo {author} {\bibfnamefont {G.}~\bibnamefont
  {Metcalfe}}, \bibinfo {author} {\bibfnamefont {I.}~\bibnamefont {Mezi{\'c}}},
  \bibinfo {author} {\bibfnamefont {A.~P.}\ \bibnamefont {{de Moura}}},
  \bibinfo {author} {\bibfnamefont {O.}~\bibnamefont {Piro}}, \bibinfo {author}
  {\bibfnamefont {M.~F.~M.}\ \bibnamefont {Speetjens}}, \bibinfo {author}
  {\bibfnamefont {R.}~\bibnamefont {Sturman}}, \bibinfo {author} {\bibfnamefont
  {J.-L.}\ \bibnamefont {Thiffeault}},\ and\ \bibinfo {author} {\bibfnamefont
  {I.}~\bibnamefont {Tuval}},\ }\bibfield  {title} {\bibinfo {title} {Frontiers
  of chaotic advection},\ }\href {https://doi.org/10.1103/RevModPhys.89.025007}
  {\bibfield  {journal} {\bibinfo  {journal} {Rev. Mod. Phys.}\ }\textbf
  {\bibinfo {volume} {89}},\ \bibinfo {pages} {025007} (\bibinfo {year}
  {2017})}\BibitemShut {NoStop}%
\bibitem [{\citenamefont {Young}(1999)}]{GFD1999}%
  \BibitemOpen
  \bibfield  {author} {\bibinfo {author} {\bibfnamefont {W.~R.}\ \bibnamefont
  {Young}},\ }\bibfield  {title} {\bibinfo {title} {Stirring and mixing},\ }in\
  \href@noop {} {\emph {\bibinfo {booktitle} {Proceedings of the 1999 Summer
  Program in Geophysical Fluid Dynamics}}},\ \bibinfo {editor} {edited by\
  \bibinfo {editor} {\bibfnamefont {J.-L.}\ \bibnamefont {Thiffeault}}\ and\
  \bibinfo {editor} {\bibfnamefont {C.}~\bibnamefont {Pasquero}}}\ (\bibinfo
  {publisher} {Woods Hole Oceanographic Institution},\ \bibinfo {address}
  {Woods Hole, MA},\ \bibinfo {year} {1999})\ \bibinfo {note} {{\tt
  http://gfd.whoi.edu/proceedings/1999/PDFvol1999.html}}\BibitemShut {NoStop}%
\bibitem [{\citenamefont {Thiffeault}(2012)}]{Thiffeault2012}%
  \BibitemOpen
  \bibfield  {author} {\bibinfo {author} {\bibfnamefont {J.-L.}\ \bibnamefont
  {Thiffeault}},\ }\bibfield  {title} {\bibinfo {title} {Using multiscale norms
  to quantify mixing and transport},\ }\href
  {https://doi.org/10.1088/0951-7715/25/2/R1} {\bibfield  {journal} {\bibinfo
  {journal} {Nonlinearity}\ }\textbf {\bibinfo {volume} {25}},\ \bibinfo
  {pages} {R1} (\bibinfo {year} {2012})},\ \Eprint
  {https://arxiv.org/abs/arXiv:1105.1101} {arXiv:1105.1101} \BibitemShut
  {NoStop}%
\bibitem [{\citenamefont {Doering}\ and\ \citenamefont
  {Nobili}(2020)}]{Doering2020}%
  \BibitemOpen
  \bibfield  {author} {\bibinfo {author} {\bibfnamefont {C.~R.}\ \bibnamefont
  {Doering}}\ and\ \bibinfo {author} {\bibfnamefont {C.}~\bibnamefont
  {Nobili}},\ }\bibinfo {title} {Lectures on stirring, mixing and transport},\
  in\ \href {https://doi.org/doi:10.1515/9783110571240-005} {\emph {\bibinfo
  {booktitle} {Transport, Fluids, and Mixing}}}\ (\bibinfo  {publisher} {De
  Gruyter Open Poland},\ \bibinfo {year} {2020})\ pp.\ \bibinfo {pages}
  {8--34}\BibitemShut {NoStop}%
\bibitem [{\citenamefont {Golestanian}\ \emph {et~al.}(2007)\citenamefont
  {Golestanian}, \citenamefont {Liverpool},\ and\ \citenamefont
  {Ajdari}}]{Golestanian2007}%
  \BibitemOpen
  \bibfield  {author} {\bibinfo {author} {\bibfnamefont {R.}~\bibnamefont
  {Golestanian}}, \bibinfo {author} {\bibfnamefont {T.~B.}\ \bibnamefont
  {Liverpool}},\ and\ \bibinfo {author} {\bibfnamefont {A.}~\bibnamefont
  {Ajdari}},\ }\bibfield  {title} {\bibinfo {title} {Designing phoretic micro-
  and nano-swimmers},\ }\href {https://doi.org/10.1088/1367-2630/9/5/126}
  {\bibfield  {journal} {\bibinfo  {journal} {New J. Phys.}\ }\textbf {\bibinfo
  {volume} {9}},\ \bibinfo {pages} {126} (\bibinfo {year} {2007})}\BibitemShut
  {NoStop}%
\bibitem [{\citenamefont {van Teeffelen}\ and\ \citenamefont
  {L\"owen}(2008)}]{vanTeeffelen2008}%
  \BibitemOpen
  \bibfield  {author} {\bibinfo {author} {\bibfnamefont {S.}~\bibnamefont {van
  Teeffelen}}\ and\ \bibinfo {author} {\bibfnamefont {H.}~\bibnamefont
  {L\"owen}},\ }\bibfield  {title} {\bibinfo {title} {Dynamics of a {B}rownian
  circle swimmer},\ }\href@noop {} {\bibfield  {journal} {\bibinfo  {journal}
  {Phys. Rev. E}\ }\textbf {\bibinfo {volume} {78}},\ \bibinfo {pages} {020101}
  (\bibinfo {year} {2008})}\BibitemShut {NoStop}%
\bibitem [{\citenamefont {Kurtzhaler}\ \emph {et~al.}(2016)\citenamefont
  {Kurtzhaler}, \citenamefont {Leitmann},\ and\ \citenamefont
  {Franosch}}]{Kurtzhaler2016}%
  \BibitemOpen
  \bibfield  {author} {\bibinfo {author} {\bibfnamefont {C.}~\bibnamefont
  {Kurtzhaler}}, \bibinfo {author} {\bibfnamefont {S.}~\bibnamefont
  {Leitmann}},\ and\ \bibinfo {author} {\bibfnamefont {T.}~\bibnamefont
  {Franosch}},\ }\bibfield  {title} {\bibinfo {title} {Intermediate scattering
  function of an anisotropic active brownian particle},\ }\href
  {https://doi.org/10.1038/srep36702} {\bibfield  {journal} {\bibinfo
  {journal} {Sci. Rep.}\ }\textbf {\bibinfo {volume} {6}},\ \bibinfo {pages}
  {36702} (\bibinfo {year} {2016})}\BibitemShut {NoStop}%
\bibitem [{\citenamefont {Lee}(2013)}]{Lee2013}%
  \BibitemOpen
  \bibfield  {author} {\bibinfo {author} {\bibfnamefont {C.~F.}\ \bibnamefont
  {Lee}},\ }\bibfield  {title} {\bibinfo {title} {Active particles under
  confinement: aggregation at the wall and gradient formation inside a
  channel},\ }\href {https://doi.org/10.1088/1367-2630/15/5/055007} {\bibfield
  {journal} {\bibinfo  {journal} {New J. Phys.}\ }\textbf {\bibinfo {volume}
  {15}},\ \bibinfo {pages} {055007} (\bibinfo {year} {2013})}\BibitemShut
  {NoStop}%
\bibitem [{\citenamefont {Ezhilan}\ and\ \citenamefont
  {Saintillan}(2015)}]{Ezhilan2015}%
  \BibitemOpen
  \bibfield  {author} {\bibinfo {author} {\bibfnamefont {B.}~\bibnamefont
  {Ezhilan}}\ and\ \bibinfo {author} {\bibfnamefont {D.}~\bibnamefont
  {Saintillan}},\ }\bibfield  {title} {\bibinfo {title} {Transport of a dilute
  active suspension in pressure-driven channel flow},\ }\href@noop {}
  {\bibfield  {journal} {\bibinfo  {journal} {J. Fluid Mech.}\ }\textbf
  {\bibinfo {volume} {777}},\ \bibinfo {pages} {482} (\bibinfo {year}
  {2015})}\BibitemShut {NoStop}%
\bibitem [{\citenamefont {Chen}\ and\ \citenamefont
  {Thiffeault}(2021)}]{ChenThiffeault2021}%
  \BibitemOpen
  \bibfield  {author} {\bibinfo {author} {\bibfnamefont {H.}~\bibnamefont
  {Chen}}\ and\ \bibinfo {author} {\bibfnamefont {J.-L.}\ \bibnamefont
  {Thiffeault}},\ }\bibfield  {title} {\bibinfo {title} {Shape matters: A
  {B}rownian microswimmer in a channel},\ }\href
  {https://doi.org/10.1017/jfm.2021.144} {\bibfield  {journal} {\bibinfo
  {journal} {J. Fluid Mech.}\ }\textbf {\bibinfo {volume} {916}},\ \bibinfo
  {pages} {A15} (\bibinfo {year} {2021})}\BibitemShut {NoStop}%
\bibitem [{\citenamefont {{D'Asaro}}\ \emph {et~al.}(2018)\citenamefont
  {{D'Asaro}}, \citenamefont {Shcherbina}, \citenamefont {Klymak},
  \citenamefont {Molemaker}, \citenamefont {Novelli}, \citenamefont {Guigand},
  \citenamefont {Haza}, \citenamefont {Haus}, \citenamefont {Ryan},
  \citenamefont {Jacobs},\ and\ \citenamefont {et~al.}}]{DAsaro2018}%
  \BibitemOpen
  \bibfield  {author} {\bibinfo {author} {\bibfnamefont {E.~A.}\ \bibnamefont
  {{D'Asaro}}}, \bibinfo {author} {\bibfnamefont {A.~Y.}\ \bibnamefont
  {Shcherbina}}, \bibinfo {author} {\bibfnamefont {J.~M.}\ \bibnamefont
  {Klymak}}, \bibinfo {author} {\bibfnamefont {J.}~\bibnamefont {Molemaker}},
  \bibinfo {author} {\bibfnamefont {G.}~\bibnamefont {Novelli}}, \bibinfo
  {author} {\bibfnamefont {C.~M.}\ \bibnamefont {Guigand}}, \bibinfo {author}
  {\bibfnamefont {A.~C.}\ \bibnamefont {Haza}}, \bibinfo {author}
  {\bibfnamefont {B.~K.}\ \bibnamefont {Haus}}, \bibinfo {author}
  {\bibfnamefont {E.~H.}\ \bibnamefont {Ryan}}, \bibinfo {author}
  {\bibfnamefont {G.~A.}\ \bibnamefont {Jacobs}},\ and\ \bibinfo {author}
  {\bibnamefont {et~al.}},\ }\bibfield  {title} {\bibinfo {title} {Ocean
  convergence and the dispersion of flotsam},\ }\href
  {https://doi.org/10.1073/pnas.1718453115} {\bibfield  {journal} {\bibinfo
  {journal} {Proc. Natl. Acad. Sci. USA}\ }\textbf {\bibinfo {volume} {115}},\
  \bibinfo {pages} {1162} (\bibinfo {year} {2018})}\BibitemShut {NoStop}%
\bibitem [{\citenamefont {Aris}(1989)}]{Aris}%
  \BibitemOpen
  \bibfield  {author} {\bibinfo {author} {\bibfnamefont {R.}~\bibnamefont
  {Aris}},\ }\href@noop {} {\emph {\bibinfo {title} {Vectors, Tensors, and the
  Basic Equations of Fluid Mechanics}}}\ (\bibinfo  {publisher} {Dover},\
  \bibinfo {address} {New York},\ \bibinfo {year} {1989})\BibitemShut {NoStop}%
\bibitem [{\citenamefont {Stone}(1990)}]{Stone1990}%
  \BibitemOpen
  \bibfield  {author} {\bibinfo {author} {\bibfnamefont {H.~A.}\ \bibnamefont
  {Stone}},\ }\bibfield  {title} {\bibinfo {title} {A simple derivation of the
  time-dependent convective-diffusion equation for surfactant transport along a
  deforming interface},\ }\href {https://doi.org/10.1063/1.857686} {\bibfield
  {journal} {\bibinfo  {journal} {Phys. Fluids A}\ }\textbf {\bibinfo {volume}
  {2}},\ \bibinfo {pages} {111–112} (\bibinfo {year} {1990})}\BibitemShut
  {NoStop}%
\bibitem [{\citenamefont {Pavliotis}(2014)}]{Pavliotis}%
  \BibitemOpen
  \bibfield  {author} {\bibinfo {author} {\bibfnamefont {G.~A.}\ \bibnamefont
  {Pavliotis}},\ }\href@noop {} {\emph {\bibinfo {title} {Stochastic Processes
  and Applications}}}\ (\bibinfo  {publisher} {Springer},\ \bibinfo {address}
  {Berlin},\ \bibinfo {year} {2014})\BibitemShut {NoStop}%
\bibitem [{\citenamefont {Risken}(1996)}]{Riksen}%
  \BibitemOpen
  \bibfield  {author} {\bibinfo {author} {\bibfnamefont {H.}~\bibnamefont
  {Risken}},\ }\href@noop {} {\emph {\bibinfo {title} {The {Fokker--Planck}
  Equation: {M}ethods of Solution and Applications}}},\ \bibinfo {edition}
  {2nd}\ ed.\ (\bibinfo  {publisher} {Springer},\ \bibinfo {address} {Berlin},\
  \bibinfo {year} {1996})\BibitemShut {NoStop}%
\bibitem [{\citenamefont {Arnold}\ \emph {et~al.}(2008)\citenamefont {Arnold},
  \citenamefont {Carlen},\ and\ \citenamefont {Ju}}]{Arnold2008}%
  \BibitemOpen
  \bibfield  {author} {\bibinfo {author} {\bibfnamefont {A.}~\bibnamefont
  {Arnold}}, \bibinfo {author} {\bibfnamefont {E.}~\bibnamefont {Carlen}},\
  and\ \bibinfo {author} {\bibfnamefont {Q.}~\bibnamefont {Ju}},\ }\bibfield
  {title} {\bibinfo {title} {Large-time behavior of non-symmetric
  {F}okker--{P}lanck type equations},\ }\href
  {https://doi.org/10.31390/cosa.2.1.11} {\bibfield  {journal} {\bibinfo
  {journal} {Comm. Stoch. Anal.}\ }\textbf {\bibinfo {volume} {2}},\ \bibinfo
  {pages} {153} (\bibinfo {year} {2008})}\BibitemShut {NoStop}%
\bibitem [{\citenamefont {Achleitner}\ \emph {et~al.}(2015)\citenamefont
  {Achleitner}, \citenamefont {Arnold},\ and\ \citenamefont
  {St\"{u}rzer}}]{Achleitner2015}%
  \BibitemOpen
  \bibfield  {author} {\bibinfo {author} {\bibfnamefont {F.}~\bibnamefont
  {Achleitner}}, \bibinfo {author} {\bibfnamefont {A.}~\bibnamefont {Arnold}},\
  and\ \bibinfo {author} {\bibfnamefont {D.}~\bibnamefont {St\"{u}rzer}},\
  }\bibfield  {title} {\bibinfo {title} {Large-time behavior in non-symmetric
  {Fokker--Planck} equations},\ }\href@noop {} {\bibfield  {journal} {\bibinfo
  {journal} {Riv. Mat. Univ. Parma}\ }\textbf {\bibinfo {volume} {6}},\
  \bibinfo {pages} {1} (\bibinfo {year} {2015})}\BibitemShut {NoStop}%
\bibitem [{\citenamefont {Arnold}\ \emph {et~al.}(2018)\citenamefont {Arnold},
  \citenamefont {Einav},\ and\ \citenamefont {W\"{o}hrer}}]{Arnold2018}%
  \BibitemOpen
  \bibfield  {author} {\bibinfo {author} {\bibfnamefont {A.}~\bibnamefont
  {Arnold}}, \bibinfo {author} {\bibfnamefont {A.}~\bibnamefont {Einav}},\ and\
  \bibinfo {author} {\bibfnamefont {T.}~\bibnamefont {W\"{o}hrer}},\ }\bibfield
   {title} {\bibinfo {title} {On the rates of decay to equilibrium in
  degenerate and defective {Fokker--Planck} equations},\ }\href
  {https://doi.org/10.1016/j.jde.2018.01.052} {\bibfield  {journal} {\bibinfo
  {journal} {J. Diff. Eqns.}\ }\textbf {\bibinfo {volume} {264}},\ \bibinfo
  {pages} {6843} (\bibinfo {year} {2018})}\BibitemShut {NoStop}%
\bibitem [{\citenamefont {Arnold}\ \emph {et~al.}(2001)\citenamefont {Arnold},
  \citenamefont {Markowich}, \citenamefont {Toscani},\ and\ \citenamefont
  {Unterreiter}}]{Arnold2001}%
  \BibitemOpen
  \bibfield  {author} {\bibinfo {author} {\bibfnamefont {A.}~\bibnamefont
  {Arnold}}, \bibinfo {author} {\bibfnamefont {P.}~\bibnamefont {Markowich}},
  \bibinfo {author} {\bibfnamefont {G.}~\bibnamefont {Toscani}},\ and\ \bibinfo
  {author} {\bibfnamefont {A.}~\bibnamefont {Unterreiter}},\ }\bibfield
  {title} {\bibinfo {title} {On convex {S}obolev inequalities and the rate of
  convergence to equilibrium for {F}okker--{P}lanck type equations},\ }\href
  {https://doi.org/10.1081/pde-100002246} {\bibfield  {journal} {\bibinfo
  {journal} {Comm. Partial Differential Equations}\ }\textbf {\bibinfo {volume}
  {26}},\ \bibinfo {pages} {43} (\bibinfo {year} {2001})}\BibitemShut {NoStop}%
\bibitem [{\citenamefont {Leli\`{e}vre}\ \emph {et~al.}(2013)\citenamefont
  {Leli\`{e}vre}, \citenamefont {Nier},\ and\ \citenamefont
  {Pavliotis}}]{Lelievre2013}%
  \BibitemOpen
  \bibfield  {author} {\bibinfo {author} {\bibfnamefont {T.}~\bibnamefont
  {Leli\`{e}vre}}, \bibinfo {author} {\bibfnamefont {F.}~\bibnamefont {Nier}},\
  and\ \bibinfo {author} {\bibfnamefont {G.~A.}\ \bibnamefont {Pavliotis}},\
  }\bibfield  {title} {\bibinfo {title} {Optimal non-reversible linear drift
  for the convergence to equilibrium of a diffusion},\ }\href
  {https://doi.org/10.1007/s10955-013-0769-x} {\bibfield  {journal} {\bibinfo
  {journal} {J. Stat. Phys.}\ }\textbf {\bibinfo {volume} {152}},\ \bibinfo
  {pages} {237} (\bibinfo {year} {2013})}\BibitemShut {NoStop}%
\bibitem [{\citenamefont {{D'Alessandro}}\ \emph {et~al.}(1999)\citenamefont
  {{D'Alessandro}}, \citenamefont {Dahleh},\ and\ \citenamefont
  {Mezi\'{c}}}]{DAlessandro1999}%
  \BibitemOpen
  \bibfield  {author} {\bibinfo {author} {\bibfnamefont {D.}~\bibnamefont
  {{D'Alessandro}}}, \bibinfo {author} {\bibfnamefont {M.}~\bibnamefont
  {Dahleh}},\ and\ \bibinfo {author} {\bibfnamefont {I.}~\bibnamefont
  {Mezi\'{c}}},\ }\bibfield  {title} {\bibinfo {title} {Control of mixing in
  fluid flow: {A} maximum entropy approach},\ }\href@noop {} {\bibfield
  {journal} {\bibinfo  {journal} {IEEE Transactions on Automatic Control}\
  }\textbf {\bibinfo {volume} {44}},\ \bibinfo {pages} {1852} (\bibinfo {year}
  {1999})}\BibitemShut {NoStop}%
\bibitem [{\citenamefont {Stremler}\ and\ \citenamefont
  {Cola}(2006)}]{Stremler2006}%
  \BibitemOpen
  \bibfield  {author} {\bibinfo {author} {\bibfnamefont {M.~A.}\ \bibnamefont
  {Stremler}}\ and\ \bibinfo {author} {\bibfnamefont {B.~A.}\ \bibnamefont
  {Cola}},\ }\bibfield  {title} {\bibinfo {title} {A maximum entropy approach
  to optimal mixing in a pulsed source-sink flow},\ }\href
  {https://doi.org/10.1063/1.2162184} {\bibfield  {journal} {\bibinfo
  {journal} {Phys. Fluids}\ }\textbf {\bibinfo {volume} {18}},\ \bibinfo
  {pages} {011701} (\bibinfo {year} {2006})}\BibitemShut {NoStop}%
\bibitem [{\citenamefont {Camesasca}\ \emph {et~al.}(2006)\citenamefont
  {Camesasca}, \citenamefont {Kaufman},\ and\ \citenamefont
  {Manas-Zloczower}}]{Camesasca2006}%
  \BibitemOpen
  \bibfield  {author} {\bibinfo {author} {\bibfnamefont {M.}~\bibnamefont
  {Camesasca}}, \bibinfo {author} {\bibfnamefont {M.}~\bibnamefont {Kaufman}},\
  and\ \bibinfo {author} {\bibfnamefont {I.}~\bibnamefont {Manas-Zloczower}},\
  }\bibfield  {title} {\bibinfo {title} {Quantifying fluid mixing with the
  {S}hannon entropy},\ }\href {https://doi.org/10.1002/mats.200600037}
  {\bibfield  {journal} {\bibinfo  {journal} {Macromolecular Theory and
  Simulations}\ }\textbf {\bibinfo {volume} {15}},\ \bibinfo {pages} {595}
  (\bibinfo {year} {2006})}\BibitemShut {NoStop}%
\bibitem [{\citenamefont {Kr\"{u}tzmann}\ \emph {et~al.}(2008)\citenamefont
  {Kr\"{u}tzmann}, \citenamefont {McDonald},\ and\ \citenamefont
  {George}}]{Krutzmann2008}%
  \BibitemOpen
  \bibfield  {author} {\bibinfo {author} {\bibfnamefont {N.~C.}\ \bibnamefont
  {Kr\"{u}tzmann}}, \bibinfo {author} {\bibfnamefont {A.~J.}\ \bibnamefont
  {McDonald}},\ and\ \bibinfo {author} {\bibfnamefont {S.~E.}\ \bibnamefont
  {George}},\ }\bibfield  {title} {\bibinfo {title} {Identification of mixing
  barriers in chemistry-climate model simulations using r\'{e}nyi entropy},\
  }\href {https://doi.org/10.1029/2007gl032829} {\bibfield  {journal} {\bibinfo
   {journal} {Geophysical Research Letters}\ }\textbf {\bibinfo {volume}
  {35}},\ \bibinfo {pages} {L06806} (\bibinfo {year} {2008})}\BibitemShut
  {NoStop}%
\bibitem [{\citenamefont {Fodor}\ and\ \citenamefont
  {Kaufman}(2011)}]{Fodor2011}%
  \BibitemOpen
  \bibfield  {author} {\bibinfo {author} {\bibfnamefont {P.~S.}\ \bibnamefont
  {Fodor}}\ and\ \bibinfo {author} {\bibfnamefont {M.}~\bibnamefont
  {Kaufman}},\ }\bibfield  {title} {\bibinfo {title} {Time evolution of mixing
  in the staggered herringbone microchannel},\ }\href
  {https://doi.org/10.1142/s0217984911026826} {\bibfield  {journal} {\bibinfo
  {journal} {Modern Physics Letters B}\ }\textbf {\bibinfo {volume} {25}},\
  \bibinfo {pages} {1111} (\bibinfo {year} {2011})}\BibitemShut {NoStop}%
\bibitem [{\citenamefont {Lauritzen}\ and\ \citenamefont
  {Thuburn}(2011)}]{Lauritzen2011}%
  \BibitemOpen
  \bibfield  {author} {\bibinfo {author} {\bibfnamefont {P.~H.}\ \bibnamefont
  {Lauritzen}}\ and\ \bibinfo {author} {\bibfnamefont {J.}~\bibnamefont
  {Thuburn}},\ }\bibfield  {title} {\bibinfo {title} {Evaluating
  advection/transport schemes using interrelated tracers, scatter plots and
  numerical mixing diagnostics},\ }\href {https://doi.org/10.1002/qj.986}
  {\bibfield  {journal} {\bibinfo  {journal} {Quarterly Journal of the Royal
  Meteorological Society}\ }\textbf {\bibinfo {volume} {138}},\ \bibinfo
  {pages} {906} (\bibinfo {year} {2011})}\BibitemShut {NoStop}%
\bibitem [{\citenamefont {Grahn}(2012)}]{Grahn2012}%
  \BibitemOpen
  \bibfield  {author} {\bibinfo {author} {\bibfnamefont {J.}~\bibnamefont
  {Grahn}},\ }\emph {\bibinfo {title} {{R}\'{e}nyi entropy and finite
  {L}yapunov exponents as metrics of transport and mixing in an idealised
  stratosphere}},\ \href@noop {} {Master's thesis},\ \bibinfo  {school}
  {Chalmers University of Technology}, \bibinfo {address} {Gothenburg, Sweden}
  (\bibinfo {year} {2012})\BibitemShut {NoStop}%
\bibitem [{\citenamefont {Brandani}\ \emph {et~al.}(2013)\citenamefont
  {Brandani}, \citenamefont {Schor}, \citenamefont {MacPhee}, \citenamefont
  {Grubmüller}, \citenamefont {Zachariae},\ and\ \citenamefont
  {Marenduzzo}}]{Brandani2013}%
  \BibitemOpen
  \bibfield  {author} {\bibinfo {author} {\bibfnamefont {G.~B.}\ \bibnamefont
  {Brandani}}, \bibinfo {author} {\bibfnamefont {M.}~\bibnamefont {Schor}},
  \bibinfo {author} {\bibfnamefont {C.~E.}\ \bibnamefont {MacPhee}}, \bibinfo
  {author} {\bibfnamefont {H.}~\bibnamefont {Grubmüller}}, \bibinfo {author}
  {\bibfnamefont {U.}~\bibnamefont {Zachariae}},\ and\ \bibinfo {author}
  {\bibfnamefont {D.}~\bibnamefont {Marenduzzo}},\ }\bibfield  {title}
  {\bibinfo {title} {Quantifying disorder through conditional entropy: An
  application to fluid mixing},\ }\href
  {https://doi.org/10.1371/journal.pone.0065617} {\bibfield  {journal}
  {\bibinfo  {journal} {PLoS ONE}\ }\textbf {\bibinfo {volume} {8}},\ \bibinfo
  {pages} {e65617} (\bibinfo {year} {2013})}\BibitemShut {NoStop}%
\bibitem [{\citenamefont {Perugini}\ \emph {et~al.}(2015)\citenamefont
  {Perugini}, \citenamefont {{De Campos}}, \citenamefont {Petrelli},
  \citenamefont {Morgavi}, \citenamefont {Vetere},\ and\ \citenamefont
  {Dingwell}}]{Perugini2015}%
  \BibitemOpen
  \bibfield  {author} {\bibinfo {author} {\bibfnamefont {D.}~\bibnamefont
  {Perugini}}, \bibinfo {author} {\bibfnamefont {C.~P.}\ \bibnamefont {{De
  Campos}}}, \bibinfo {author} {\bibfnamefont {M.}~\bibnamefont {Petrelli}},
  \bibinfo {author} {\bibfnamefont {D.}~\bibnamefont {Morgavi}}, \bibinfo
  {author} {\bibfnamefont {F.~P.}\ \bibnamefont {Vetere}},\ and\ \bibinfo
  {author} {\bibfnamefont {D.~B.}\ \bibnamefont {Dingwell}},\ }\bibfield
  {title} {\bibinfo {title} {Quantifying magma mixing with the {S}hannon
  entropy: Application to simulations and experiments},\ }\href
  {https://doi.org/10.1016/j.lithos.2015.09.008} {\bibfield  {journal}
  {\bibinfo  {journal} {Lithos}\ }\textbf {\bibinfo {volume} {236-237}},\
  \bibinfo {pages} {299} (\bibinfo {year} {2015})}\BibitemShut {NoStop}%
\bibitem [{\citenamefont {Boyland}\ \emph {et~al.}(2000)\citenamefont
  {Boyland}, \citenamefont {Aref},\ and\ \citenamefont
  {Stremler}}]{Boyland2000}%
  \BibitemOpen
  \bibfield  {author} {\bibinfo {author} {\bibfnamefont {P.~L.}\ \bibnamefont
  {Boyland}}, \bibinfo {author} {\bibfnamefont {H.}~\bibnamefont {Aref}},\ and\
  \bibinfo {author} {\bibfnamefont {M.~A.}\ \bibnamefont {Stremler}},\
  }\bibfield  {title} {\bibinfo {title} {Topological fluid mechanics of
  stirring},\ }\href@noop {} {\bibfield  {journal} {\bibinfo  {journal} {J.
  Fluid Mech.}\ }\textbf {\bibinfo {volume} {403}},\ \bibinfo {pages} {277}
  (\bibinfo {year} {2000})}\BibitemShut {NoStop}%
\bibitem [{\citenamefont {Thiffeault}\ and\ \citenamefont
  {Finn}(2006)}]{Thiffeault2006}%
  \BibitemOpen
  \bibfield  {author} {\bibinfo {author} {\bibfnamefont {J.-L.}\ \bibnamefont
  {Thiffeault}}\ and\ \bibinfo {author} {\bibfnamefont {M.~D.}\ \bibnamefont
  {Finn}},\ }\bibfield  {title} {\bibinfo {title} {Topology, braids, and mixing
  in fluids},\ }\href {https://doi.org/10.1098/rsta.2006.1899} {\bibfield
  {journal} {\bibinfo  {journal} {Philos. Trans. Royal Soc. Lond. A}\ }\textbf
  {\bibinfo {volume} {364}},\ \bibinfo {pages} {3251} (\bibinfo {year}
  {2006})}\BibitemShut {NoStop}%
\bibitem [{\citenamefont {Gouillart}\ \emph {et~al.}(2006)\citenamefont
  {Gouillart}, \citenamefont {Finn},\ and\ \citenamefont
  {Thiffeault}}]{Gouillart2006}%
  \BibitemOpen
  \bibfield  {author} {\bibinfo {author} {\bibfnamefont {E.}~\bibnamefont
  {Gouillart}}, \bibinfo {author} {\bibfnamefont {M.~D.}\ \bibnamefont
  {Finn}},\ and\ \bibinfo {author} {\bibfnamefont {J.-L.}\ \bibnamefont
  {Thiffeault}},\ }\bibfield  {title} {\bibinfo {title} {Topological mixing
  with ghost rods},\ }\href@noop {} {\bibfield  {journal} {\bibinfo  {journal}
  {Phys. Rev. E}\ }\textbf {\bibinfo {volume} {73}},\ \bibinfo {pages} {036311}
  (\bibinfo {year} {2006})}\BibitemShut {NoStop}%
\bibitem [{\citenamefont {\"{O}sterreicher}\ and\ \citenamefont
  {Vajda}(2003)}]{Osterreicher2003}%
  \BibitemOpen
  \bibfield  {author} {\bibinfo {author} {\bibfnamefont {F.}~\bibnamefont
  {\"{O}sterreicher}}\ and\ \bibinfo {author} {\bibfnamefont {I.}~\bibnamefont
  {Vajda}},\ }\bibfield  {title} {\bibinfo {title} {A new class of metric
  divergences on probability spaces and its applicability in statistics},\
  }\href {https://doi.org/10.1007/bf02517812} {\bibfield  {journal} {\bibinfo
  {journal} {Annals of the Institute of Statistical Mathematics}\ }\textbf
  {\bibinfo {volume} {55}},\ \bibinfo {pages} {639} (\bibinfo {year}
  {2003})}\BibitemShut {NoStop}%
\bibitem [{\citenamefont {Liese}\ and\ \citenamefont
  {Vajda}(2006)}]{Liese2006}%
  \BibitemOpen
  \bibfield  {author} {\bibinfo {author} {\bibfnamefont {F.}~\bibnamefont
  {Liese}}\ and\ \bibinfo {author} {\bibfnamefont {I.}~\bibnamefont {Vajda}},\
  }\bibfield  {title} {\bibinfo {title} {On divergences and informations in
  statistics and information theory},\ }\href
  {https://doi.org/10.1109/tit.2006.881731} {\bibfield  {journal} {\bibinfo
  {journal} {IEEE Transactions on Information Theory}\ }\textbf {\bibinfo
  {volume} {52}},\ \bibinfo {pages} {4394–4412} (\bibinfo {year}
  {2006})}\BibitemShut {NoStop}%
\bibitem [{\citenamefont {Cover}\ and\ \citenamefont
  {Thomas}(2005)}]{CoverThomas}%
  \BibitemOpen
  \bibfield  {author} {\bibinfo {author} {\bibfnamefont {T.~M.}\ \bibnamefont
  {Cover}}\ and\ \bibinfo {author} {\bibfnamefont {J.~A.}\ \bibnamefont
  {Thomas}},\ }\href@noop {} {\emph {\bibinfo {title} {Elements of information
  theory}}},\ \bibinfo {edition} {2nd}\ ed.\ (\bibinfo  {publisher} {Wiley},\
  \bibinfo {address} {Hoboken, New Jersey},\ \bibinfo {year}
  {2005})\BibitemShut {NoStop}%
\bibitem [{\citenamefont {Endres}\ and\ \citenamefont
  {Schindelin}(2003)}]{Endres2003}%
  \BibitemOpen
  \bibfield  {author} {\bibinfo {author} {\bibfnamefont {D.~M.}\ \bibnamefont
  {Endres}}\ and\ \bibinfo {author} {\bibfnamefont {J.~E.}\ \bibnamefont
  {Schindelin}},\ }\bibfield  {title} {\bibinfo {title} {A new metric for
  probability distributions},\ }\href {https://doi.org/10.1109/tit.2003.813506}
  {\bibfield  {journal} {\bibinfo  {journal} {IEEE Transactions on Information
  Theory}\ }\textbf {\bibinfo {volume} {49}},\ \bibinfo {pages} {1858}
  (\bibinfo {year} {2003})}\BibitemShut {NoStop}%
\bibitem [{\citenamefont {Pierrehumbert}(1994)}]{Pierrehumbert1994}%
  \BibitemOpen
  \bibfield  {author} {\bibinfo {author} {\bibfnamefont {R.~T.}\ \bibnamefont
  {Pierrehumbert}},\ }\bibfield  {title} {\bibinfo {title} {Tracer
  microstructure in the large-eddy dominated regime},\ }\href@noop {}
  {\bibfield  {journal} {\bibinfo  {journal} {Chaos Solitons Fractals}\
  }\textbf {\bibinfo {volume} {4}},\ \bibinfo {pages} {1091} (\bibinfo {year}
  {1994})}\BibitemShut {NoStop}%
\bibitem [{\citenamefont {H\"{o}rmander}(1990)}]{HormanderI}%
  \BibitemOpen
  \bibfield  {author} {\bibinfo {author} {\bibfnamefont {L.}~\bibnamefont
  {H\"{o}rmander}},\ }\href@noop {} {\emph {\bibinfo {title} {The Analysis of
  Linear Partial Differential Operators}}},\ \bibinfo {edition} {2nd}\ ed.,\
  Vol.~\bibinfo {volume} {1}\ (\bibinfo  {publisher} {Springer},\ \bibinfo
  {address} {Berlin},\ \bibinfo {year} {1990})\BibitemShut {NoStop}%
\bibitem [{\citenamefont {Nakamura}(1996)}]{Nakamura1996}%
  \BibitemOpen
  \bibfield  {author} {\bibinfo {author} {\bibfnamefont {N.}~\bibnamefont
  {Nakamura}},\ }\bibfield  {title} {\bibinfo {title} {Two-dimensional mixing,
  edge formation, and permeability diagnosed in an area coordinate},\
  }\href@noop {} {\bibfield  {journal} {\bibinfo  {journal} {J. Atmos. Sci.}\
  }\textbf {\bibinfo {volume} {53}},\ \bibinfo {pages} {1524} (\bibinfo {year}
  {1996})}\BibitemShut {NoStop}%
\bibitem [{\citenamefont {Mathew}\ \emph {et~al.}(2003)\citenamefont {Mathew},
  \citenamefont {Mezi{\'c}},\ and\ \citenamefont {Petzold}}]{Mathew2003}%
  \BibitemOpen
  \bibfield  {author} {\bibinfo {author} {\bibfnamefont {G.}~\bibnamefont
  {Mathew}}, \bibinfo {author} {\bibfnamefont {I.}~\bibnamefont {Mezi{\'c}}},\
  and\ \bibinfo {author} {\bibfnamefont {L.}~\bibnamefont {Petzold}},\
  }\bibfield  {title} {\bibinfo {title} {A multiscale measure for mixing and
  its applications},\ }in\ \href@noop {} {\emph {\bibinfo {booktitle} {Proc.
  Conf. on Decision and Control, Maui, HI}}},\ \bibinfo {organization} {IEEE}\
  (\bibinfo  {publisher} {IEEE},\ \bibinfo {year} {2003})\BibitemShut {NoStop}%
\bibitem [{\citenamefont {Mathew}\ \emph {et~al.}(2005)\citenamefont {Mathew},
  \citenamefont {Mezi{\'c}},\ and\ \citenamefont {Petzold}}]{Mathew2005}%
  \BibitemOpen
  \bibfield  {author} {\bibinfo {author} {\bibfnamefont {G.}~\bibnamefont
  {Mathew}}, \bibinfo {author} {\bibfnamefont {I.}~\bibnamefont {Mezi{\'c}}},\
  and\ \bibinfo {author} {\bibfnamefont {L.}~\bibnamefont {Petzold}},\
  }\bibfield  {title} {\bibinfo {title} {A multiscale measure for mixing},\
  }\href@noop {} {\bibfield  {journal} {\bibinfo  {journal} {Physica D}\
  }\textbf {\bibinfo {volume} {211}},\ \bibinfo {pages} {23} (\bibinfo {year}
  {2005})}\BibitemShut {NoStop}%
\bibitem [{\citenamefont {Mathew}\ \emph {et~al.}(2007)\citenamefont {Mathew},
  \citenamefont {Mezi{\'c}}, \citenamefont {Grivopoulos}, \citenamefont
  {Vaidya},\ and\ \citenamefont {Petzold}}]{Mathew2007}%
  \BibitemOpen
  \bibfield  {author} {\bibinfo {author} {\bibfnamefont {G.}~\bibnamefont
  {Mathew}}, \bibinfo {author} {\bibfnamefont {I.}~\bibnamefont {Mezi{\'c}}},
  \bibinfo {author} {\bibfnamefont {S.}~\bibnamefont {Grivopoulos}}, \bibinfo
  {author} {\bibfnamefont {U.}~\bibnamefont {Vaidya}},\ and\ \bibinfo {author}
  {\bibfnamefont {L.}~\bibnamefont {Petzold}},\ }\bibfield  {title} {\bibinfo
  {title} {Optimal control of mixing in {S}tokes fluid flows},\ }\href@noop {}
  {\bibfield  {journal} {\bibinfo  {journal} {J. Fluid Mech.}\ }\textbf
  {\bibinfo {volume} {580}},\ \bibinfo {pages} {261} (\bibinfo {year}
  {2007})}\BibitemShut {NoStop}%
\bibitem [{\citenamefont {Lin}\ \emph {et~al.}(2011)\citenamefont {Lin},
  \citenamefont {Doering},\ and\ \citenamefont {Thiffeault}}]{Lin2011b}%
  \BibitemOpen
  \bibfield  {author} {\bibinfo {author} {\bibfnamefont {Z.}~\bibnamefont
  {Lin}}, \bibinfo {author} {\bibfnamefont {C.~R.}\ \bibnamefont {Doering}},\
  and\ \bibinfo {author} {\bibfnamefont {J.-L.}\ \bibnamefont {Thiffeault}},\
  }\bibfield  {title} {\bibinfo {title} {Optimal stirring strategies for
  passive scalar mixing},\ }\href {https://doi.org/10.1017/S0022112011000292}
  {\bibfield  {journal} {\bibinfo  {journal} {J. Fluid Mech.}\ }\textbf
  {\bibinfo {volume} {675}},\ \bibinfo {pages} {465} (\bibinfo {year}
  {2011})}\BibitemShut {NoStop}%
\bibitem [{\citenamefont {Foures}\ \emph {et~al.}(2014)\citenamefont {Foures},
  \citenamefont {Caulfield},\ and\ \citenamefont {Schmid}}]{Foures2014}%
  \BibitemOpen
  \bibfield  {author} {\bibinfo {author} {\bibfnamefont {D.~P.~G.}\
  \bibnamefont {Foures}}, \bibinfo {author} {\bibfnamefont {C.~P.}\
  \bibnamefont {Caulfield}},\ and\ \bibinfo {author} {\bibfnamefont {P.~J.}\
  \bibnamefont {Schmid}},\ }\bibfield  {title} {\bibinfo {title} {Optimal
  mixing in two-dimensional plane {P}oiseuille flow at finite p\'{e}clet
  number},\ }\href {https://doi.org/10.1017/jfm.2014.182} {\bibfield  {journal}
  {\bibinfo  {journal} {J. Fluid Mech.}\ }\textbf {\bibinfo {volume} {748}},\
  \bibinfo {pages} {241} (\bibinfo {year} {2014})}\BibitemShut {NoStop}%
\bibitem [{\citenamefont {Vermach}\ and\ \citenamefont
  {Caulfield}(2018)}]{Vermach2018}%
  \BibitemOpen
  \bibfield  {author} {\bibinfo {author} {\bibfnamefont {L.}~\bibnamefont
  {Vermach}}\ and\ \bibinfo {author} {\bibfnamefont {C.~P.}\ \bibnamefont
  {Caulfield}},\ }\bibfield  {title} {\bibinfo {title} {Optimal mixing in
  three-dimensional plane {P}oiseuille flow at high {P\'eclet} number},\
  }\href@noop {} {\bibfield  {journal} {\bibinfo  {journal} {J. Fluid Mech.}\
  }\textbf {\bibinfo {volume} {850}},\ \bibinfo {pages} {875} (\bibinfo {year}
  {2018})}\BibitemShut {NoStop}%
\bibitem [{\citenamefont {Marcotte}\ and\ \citenamefont
  {Caulfield}(2018)}]{Marcotte2018b}%
  \BibitemOpen
  \bibfield  {author} {\bibinfo {author} {\bibfnamefont {F.}~\bibnamefont
  {Marcotte}}\ and\ \bibinfo {author} {\bibfnamefont {C.~P.}\ \bibnamefont
  {Caulfield}},\ }\bibfield  {title} {\bibinfo {title} {Optimal mixing in {2D}
  stratified plane {P}oiseuille flow at finite {P}\'{e}clet and {R}ichardson
  numbers},\ }\href@noop {} {\bibfield  {journal} {\bibinfo  {journal} {J.
  Fluid Mech.}\ }\textbf {\bibinfo {volume} {853}},\ \bibinfo {pages} {359}
  (\bibinfo {year} {2018})}\BibitemShut {NoStop}%
\bibitem [{\citenamefont {Thiffeault}\ \emph {et~al.}(2004)\citenamefont
  {Thiffeault}, \citenamefont {Doering},\ and\ \citenamefont
  {Gibbon}}]{Thiffeault2004}%
  \BibitemOpen
  \bibfield  {author} {\bibinfo {author} {\bibfnamefont {J.-L.}\ \bibnamefont
  {Thiffeault}}, \bibinfo {author} {\bibfnamefont {C.~R.}\ \bibnamefont
  {Doering}},\ and\ \bibinfo {author} {\bibfnamefont {J.~D.}\ \bibnamefont
  {Gibbon}},\ }\bibfield  {title} {\bibinfo {title} {A bound on mixing
  efficiency for the advection--diffusion equation},\ }\href@noop {} {\bibfield
   {journal} {\bibinfo  {journal} {J. Fluid Mech.}\ }\textbf {\bibinfo {volume}
  {521}},\ \bibinfo {pages} {105} (\bibinfo {year} {2004})}\BibitemShut
  {NoStop}%
\bibitem [{\citenamefont {Doering}\ and\ \citenamefont
  {Thiffeault}(2006)}]{DoeringThiffeault2006}%
  \BibitemOpen
  \bibfield  {author} {\bibinfo {author} {\bibfnamefont {C.~R.}\ \bibnamefont
  {Doering}}\ and\ \bibinfo {author} {\bibfnamefont {J.-L.}\ \bibnamefont
  {Thiffeault}},\ }\bibfield  {title} {\bibinfo {title} {Multiscale mixing
  efficiencies for steady sources},\ }\href@noop {} {\bibfield  {journal}
  {\bibinfo  {journal} {Phys. Rev. E}\ }\textbf {\bibinfo {volume} {74}},\
  \bibinfo {pages} {025301(R)} (\bibinfo {year} {2006})}\BibitemShut {NoStop}%
\bibitem [{\citenamefont {Shaw}\ \emph {et~al.}(2007)\citenamefont {Shaw},
  \citenamefont {Thiffeault},\ and\ \citenamefont {Doering}}]{Shaw2007}%
  \BibitemOpen
  \bibfield  {author} {\bibinfo {author} {\bibfnamefont {T.~A.}\ \bibnamefont
  {Shaw}}, \bibinfo {author} {\bibfnamefont {J.-L.}\ \bibnamefont
  {Thiffeault}},\ and\ \bibinfo {author} {\bibfnamefont {C.~R.}\ \bibnamefont
  {Doering}},\ }\bibfield  {title} {\bibinfo {title} {Stirring up trouble:
  {M}ulti-scale mixing measures for steady scalar sources},\ }\href@noop {}
  {\bibfield  {journal} {\bibinfo  {journal} {Physica D}\ }\textbf {\bibinfo
  {volume} {231}},\ \bibinfo {pages} {143} (\bibinfo {year} {2007})},\ \Eprint
  {https://arxiv.org/abs/arXiv:physics/0607270} {arXiv:physics/0607270}
  \BibitemShut {NoStop}%
\bibitem [{\citenamefont {Thiffeault}\ and\ \citenamefont
  {Pavliotis}(2008)}]{Thiffeault2008}%
  \BibitemOpen
  \bibfield  {author} {\bibinfo {author} {\bibfnamefont {J.-L.}\ \bibnamefont
  {Thiffeault}}\ and\ \bibinfo {author} {\bibfnamefont {G.~A.}\ \bibnamefont
  {Pavliotis}},\ }\bibfield  {title} {\bibinfo {title} {Optimizing the source
  distribution in fluid mixing},\ }\href
  {https://doi.org/10.1016/j.physd.2007.11.013} {\bibfield  {journal} {\bibinfo
   {journal} {Physica D}\ }\textbf {\bibinfo {volume} {237}},\ \bibinfo {pages}
  {918} (\bibinfo {year} {2008})},\ \Eprint
  {https://arxiv.org/abs/arXiv:physics/0703135} {arXiv:physics/0703135}
  \BibitemShut {NoStop}%
\end{thebibliography}%

\appendix

\section{Derivation of~\texorpdfstring{\cref{eq:fdiv_dot}}{(IV.4)}}
\label{apx:fdiv_dot}

Write~$\pd_\t\pp_i = -\grad\cdot\Fv(\pp_i)$,
with~$\Fv(\pp) \ldef \uv\,\pp - \Dbb\cdot\grad\pp$, and consider a
domain~$\Omega \subset \mathbb{R}^n$ with~$\Fv(\pp_i)\cdot\nuv=0$
on~$\pd\Omega$.  Then by direct differentiation of \cref{eq:fdiv}:
\begin{align*}
  \skew{-3}\dot\Df[\pp_1,\pp_2]
  &=
  \int_\Omega
  \l(
  \pd_\t\pp_2\,\ffc(\pp_1/\pp_2)
  +
  \pp_2\,\ffc'(\pp_1/\pp_2)
  \l(\pd_\t\pp_1/\pp_2 - \pp_1\,\pd_\t\pp_2/\pp_2^2\r)
  \r)\dV
  \nonumber\\
  &=
  -
  \int_\Omega
  \l(
  \grad\cdot\Fv(\pp_2)\,\ffc(\pp_1/\pp_2)
  +
  \ffc'(\pp_1/\pp_2)
  \l(\grad\cdot\Fv(\pp_1)
  - (\pp_1/\pp_2)\grad\cdot\Fv(\pp_2)\r)
  \r)\dV.
\end{align*}
We integrate by parts, and two terms
containing~$\ffc'(\pp_1/\pp_2)\,\Fv(\pp_2)\cdot\grad(\pp_1/\pp_2)$ cancel.  We
are left with
\begin{equation}
  \dot\Df[\pp_1,\pp_2]
  =
  \text{BT}[\pp_1,\pp_2]
  +
  \int_\Omega
  \pp_2^{-1}\ffc''(\pp_1/\pp_2)\grad(\pp_1/\pp_2)\cdot
  \l(\pp_2\Fv(\pp_1) - \pp_1\Fv(\pp_2)\r)\dV
  \label{eq:IBP_BT}
\end{equation}
with the boundary terms
\begin{align*}
  \text{BT}[\pp_1,\pp_2]
  &=
  -
  \int_{\pd\Omega}
  \ffc(\pp_1/\pp_2)\,\Fv(\pp_2)\cdot\nuv
  +
  \pp_2^{-1}\,\ffc'(\pp_1/\pp_2)
  \l(\pp_2\Fv(\pp_1) - \pp_1\Fv(\pp_2)\r)\cdot\dSv.
\end{align*}
The boundary terms vanish when~$\Fv(\pp_i)\cdot\nuv=0$ on~$\pd\Omega$.
Also,
\begin{align}
  \pp_2\Fv(\pp_1) - \pp_1\Fv(\pp_2)
  &=
  \pp_2(\uv\,\pp_1 - \Dbb\cdot\grad\pp_1)
  -
  \pp_1(\uv\,\pp_2 - \Dbb\cdot\grad\pp_2)
  \nonumber\\
  &=
  -\pp_2\Dbb\cdot\grad\pp_1
  + \pp_1\Dbb\cdot\grad\pp_2
  \nonumber\\
  &=
  -\pp_2^2\,\Dbb\cdot\grad(\pp_1/\pp_2).
  \label{eq:p2F1_p1F2}
\end{align}
Inserting~\eqref{eq:p2F1_p1F2} into~\eqref{eq:IBP_BT} recovers
\cref{eq:fdiv_dot}.

\section{Decay of~\texorpdfstring{$L^\qn$}{Lq} norms}
\label{apx:Lq_decay}

For~$1 \le \qn \le \infty$, does any~$L^\qn$ norm other than~$\qn=1$ decay
monotonically?  If we put~$G(\theta) = \lvert\theta\rvert^\qn$ in
\cref{eq:dt_intG}, then for~$\qn$ even we have~$G(\theta)=\theta^\qn$,
$G'(\theta)=\qn\theta^{\qn-1}$, and $G''(\theta) = \qn(\qn-1)\theta^{\qn-2}$,
and find
\begin{equation}
  \frac{\d}{\dt}\lVert\theta\rVert_\qn^\qn
  =
  \qn(\qn-1)\int_\Omega
  \theta^{\qn-1}\,\uv\cdot\grad\theta\dV
  -
  \qn(\qn-1)\int_\Omega
  \theta^{\qn-2}\,\grad\theta\cdot\Dbb\cdot\grad\theta\dV
\end{equation}
where we set~$\K=\k=0$ for convenience.  The first term on the right is not
sign definite for any even~$\qn$, so none of these norms will necessarily
decay monotonically.  For~$\qn$ odd, we have
that~$G'(\theta) = \qn\theta^{\qn-1}\sgn(\theta)$
and~$G''(\theta) = 2\qn\theta^{\qn-1}\delta(\theta) +
\qn(\qn-1)\theta^{\qn-2}\sgn(\theta)$, and \cref{eq:dt_intG} becomes
\begin{multline}
  \frac{\d}{\dt}\lVert\theta\rVert_\qn^\qn
  =
  \qn(\qn-1)\int_\Omega
  \theta^{\qn-2}\lvert\theta\rvert\,\uv\cdot\grad\theta\dV \\
  -
  2\qn\int_\Omega \theta^{\qn-1}\delta(\theta)
  \,\grad\theta\cdot\Dbb\cdot\grad\theta\dV
  -
  \qn(\qn-1)\int_\Omega \lvert\theta\rvert^{\qn-2}
  \,\grad\theta\cdot\Dbb\cdot\grad\theta\dV.
\end{multline}
For~$\qn=1$ this reduces to \cref{eq:L1_evol}; for~$\qn>1$ we have
\begin{equation}
  \frac{\d}{\dt}\lVert\theta\rVert_\qn^\qn
  =
  \qn(\qn-1)\int_\Omega
  \theta^{\qn-2}\lvert\theta\rvert\,\uv\cdot\grad\theta\dV
  -
  \qn(\qn-1)\int_\Omega \lvert\theta\rvert^{\qn-2}
  \,\grad\theta\cdot\Dbb\cdot\grad\theta\dV.
\end{equation}
and again the first term is not sign-definite.  We conclude that the~$L^1$
norm is the only such norm that exhibits a monotonic decay to zero in the
nonuniform mixing case.

\printnomenclature

\end{document}